\documentclass[apj]{emulateapj} 
 \usepackage{fancyhdr, mathtools,adjustbox,geometry, graphicx, epstopdf,subfigure, amssymb, amsmath,amstext,natbib,titlesec, longtable,multirow, alltt, indentfirst, ifthen, hyperref,wrapfig}
\newcommand{\cm}{\text{cm}}

\newcommand{\AU}{\text{AU}}

\newcommand{\m}{\text{m}}

\newcommand{\g}{\text{g}}

\titleformat{\section}{\normalfont\center\scshape}{\thesection.}{1em}{}

\begin{document}

\bibliographystyle{apj}

\slugcomment{Accepted for Publication in ApJ: October 21, 2015}

\title{Resolved Millimeter-Wavelength Observations of Debris Disks around Solar-Type Stars}

\author{Amy Steele\altaffilmark{1,2}, A. Meredith Hughes\altaffilmark{1}, John Carpenter
\altaffilmark{3}, Angelo Ricarte\altaffilmark{4}, Sean M. Andrews\altaffilmark{5}, David J. Wilner\altaffilmark{5}, Eugene Chiang\altaffilmark{6}}
\altaffiltext{1}{Department of Astronomy, Van Vleck Observatory, Wesleyan University, 96 Foss Hill Drive, Middletown, CT, 06459, USA }
\altaffiltext{2}{Department of Astronomy, 1113 Physical Sciences Complex, Bldg. 415, University of Maryland, College Park, MD 20742-2421, USA; {\tt asteele@astro.umd.edu}}
\altaffiltext{3}{Division of Physics, Mathematics, \& Astronomy, MC249-17, California Institute of Technology, Pasadena, CA 91125, USA}
\altaffiltext{4}{J. W. Gibbs Laboratory, Department of Astronomy, Yale University, 260 Whitney Avenue, New Haven, CT 06511, USA}
\altaffiltext{5}{Harvard-Smithsonian Center for Astrophysics, MS-42, 60 Garden Street, Cambridge, MA 02138, USA}
\altaffiltext{6}{Department of Astronomy, 501 Campbell Hall, University of California, Berkeley, CA 94720-3411, USA}

 \begin{abstract}
 The presence of debris disks around young main sequence stars hints at the existence and structure of planetary systems. Millimeter-wavelength observations probe large grains that trace the location of planetesimal belts. The FEPS (Formation and Evolution of Planetary Systems) \textit{Spitzer} Legacy survey of nearby young solar analogues yielded a sample of five debris disk-hosting stars with millimeter flux suitable for interferometric follow-up. We present observations with the Submillimeter Array (SMA) and the Combined Array for Research in Millimeter-wave Astronomy (CARMA) at $\sim$2'' resolution that spatially resolve the debris disks around these nearby ($d\sim$50 pc) stars. Two of the five disks (HD 377, HD 8907) are spatially resolved for the first time and one (HD 104860) is resolved at millimeter wavelengths for the first time. We combine our new observations with archival SMA and Atacama Large Millimeter/Submillimeter Array (ALMA) data to enable a uniform analysis of the full five-object sample.  We simultaneously model the broad-band photometric data and resolved millimeter visibilities to constrain the dust temperatures and disk morphologies, and perform an MCMC analysis to fit for basic structural parameters. We find that the radii and widths of the cold outer belts exhibit properties consistent with scaled-up versions of the Solar System's Kuiper Belt. All the disks exhibit characteristic grain sizes comparable to the blowout size, and all the resolved observations of emission from large dust grains are consistent with an axisymmetric dust distribution to within the uncertainties. These results are consistent with comparable studies carried out at infrared wavelengths.
 \end{abstract}

\maketitle
\section{Introduction}

Planets form in circumstellar disks as a natural by-product of star formation. From observations with the {\it Kepler} telescope, we now know that planetary systems are a common outcome of the planet formation process with an average frequency of at least one planet per star \citep{Swift}.  At the relatively small semimajor axis range probed by Kepler ($\lesssim 1$\,AU) the properties of these systems differ dramatically from those of our solar system, but far less is known about the properties of planetary systems at large semimajor axis, particularly Uranus and Neptune analogs. Optically thin, second-generation debris disks are similarly a very common phenomenon, occurring around $\sim\!17\%$ of FGK stars (DEBRIS survey, \citealt{Matthews14}). The existence of debris disks around a large fraction of solar-type stars suggests that leftover planetesimal belts analogous to the asteroid and comet reservoirs of the solar system are common \citep{MoroM13}.  Observations that spatially resolve the radial distribution of dust around these stars provide insight into the semimajor axis distribution of Kuiper Belt analogs, and can hint at the properties of wide-separation planets that may be sculpting the belts.

Millimeter-wavelength interferometry provides sufficient angular resolution to reveal the spatial distribution of large dust grains that are not significantly affected by radiation pressure.  Millimeter-wavelength observations are highly complementary to observations at shorter wavelengths that probe the spatial distribution of smaller dust grains that are subject to different physical processes.  At the time of submission, eight debris disks have been spatially resolved using (sub)millimeter interferometry: Fomalhaut \citep{Boley12}, HR 8799 \citep{Hughes11}, AU Mic (\citealt{Wilner12,MacGregor13}), $\beta$ Pictoris (\citealt{Wilner11,Dent14}), HD 107146 (\citealt{Corder09,Hughes11,Ricci15}), HD 61005 \citep{Ricarte13}, HD 32297 \citep{Maness08}, and HD 21997 (\citealt{Kospal13,Moor13}).  These observations have revealed a striking diversity of debris disk properties: narrow millimeter rings with or without scattered light haloes, broad rings with or without haloes, unusually large quantities of molecular gas, and evidence for interactions with the interstellar medium.  Each uses a different analysis technique, and the wide range of stellar masses, distances, and angular resolutions represented by these observations makes it difficult to draw general conclusions about the spatial distributions of large grains around nearby stars. 

In this paper, we assemble a uniform sample of debris disks around solar-type stars observed using millimeter interferometry and analyzed in a consistent manner to allow for robust comparisons of debris disk properties. The five disks in this sample (HD 377, HD 8907, HD 61005, HD 104860, and HD 107146) were drawn from the Formation and Evolution of Planetary Systems (FEPS) Spitzer Legacy Science Program. The survey made extensive and uniform IR observations of 328 nearby stars with ages ranging from 3 Myr to 3 Gyr and masses within $0.8-1.5\,M_\odot$, (95\% of the sample), with the goal of detecting disks at varying stages of evolution. The program utilized all three Spitzer science instruments, IRAC (Infrared Array Camera), IRS (Infrared Spectrograph), and MIPS (Multiband Imaging Photometer), for all stars in the sample to collect photometry and spectroscopy at 24, 70, and 160 $\mu$m to infer the approximate radial distribution of dust in these systems. They detected continuum excess around 54 sources at either 24 or 33 $\mu$m, which \cite{Carpenter08} characterized with blackbody fitting. In addition to providing a uniform sample of stars, photometry provided by the FEPS survey is well-sampled. The five sources in this work are the only debris disk sources that were detected  in long-wavelength follow-up observations by \citet{Rocca09} at wavelengths of both 350\,$\mu$m and 1.2\,mm. \cite{Rocca09} present a summary of previous studies of the disks in this sample, with \cite{Ertel11} expanding upon the studies of HD 107146. The basic properties of the sample are given in Table 1.

Here we present a set of submillimeter-wavelength observations of the five disks in our sample (Section 2). HD 377 and HD 8907 are resolved for the first time and HD 104860 is resolved at millimeter wavelengths for the first time. We simultaneously model the excess flux from the spectral energy distribution (SED) and spatially resolved visibilities, and fit for basic physical characteristics of the disks (mass, characteristic grain size, inner radius, belt width, and long-wavelength slope of the dust opacity, $\beta$), with a uniform analysis. We describe the modeling process in Section 3, and present the results of the analysis in Section 4. We focus on the following major topics: (1) By spatially resolving the disks, we unambiguously determine the radius of the debris belts. We can use this spatial information to derive basic properties of the dust size distribution. (2) The spatially resolved data allows us to place constraints on the radial width of the debris belts, which can be compared with observations at other wavelengths to provide insight into the physics of collisional particle grinding and radiation pressure effects. (3) Finally, we search for deviations from axisymmetry, which may be caused by the dynamical influence of unseen Uranus and Neptune analogs in the disk.  We discuss the implications of our results and place our observations in context with multiwavelength debris disk studies (Section 5).

\section{Observations}
\label{sec:observations}

We selected targets for observation in order to complete the sample of five debris disk-hosting stars detected by the FEPS survey at both 350\,$\mu$m and 1.2\,mm wavelengths (HD 191089 was excluded from the sample due to a non-detection at 1.2\,mm). The resulting sample of five systems spans stellar ages from $\sim\!30-300$\,Myr, spectral types from F8 to G3/5V, and luminosities from $0.6-2\,L_\sun$. Properties of the host stars for each of the observed disks are listed in Table~\ref{tab:starprops}. Each of the disks in this sample was observed with the Submillimeter Array (SMA). Additional observations were made with the Combined Array for Millimeter-wave Astronomy (CARMA) and the Atacama Large Millimeter/sub-millimeter Array (ALMA) for a subset of the sample. The disks were observed at 230\,GHz (1.3\,mm), and/or 345\,GHz (0.87\,mm). The frequency and antenna configurations for each disk was chosen to best match the predicted disk size based on SED modeling. 
\begin{center}
\begin{table*}
\caption{Stellar Properties}
\resizebox{\textwidth}{!}{
\begin{tabular}{cccccccc}
\hline\hline\noalign{\smallskip}\noalign{\smallskip}
Source & $d$ (pc) & Age (Myr)& RA (J2000) &Dec (J2000) & Spectral Type & $T_\text{eff}$ (K)& $\log(L/L_\odot)$ \\
\noalign{\smallskip}\hline\noalign{\smallskip}\noalign{\smallskip}
HD 377 & 40 & 32 & 00:08:26 & +06:37:01 &G2 V & 5852 & 0.09 \\
HD 8907 & 34 & 320 & 01:28:34 & +42:16:04 & F8 & 6250 & 0.32 \\
HD 61005 & 35 & 100 &07:35:47 &$-3$2:12:14  & G3/5 V & 5456 & $-0.25$ \\
HD 104860 & 48 & 32 &12:04:34 & +66:20:12   & F8 & 5950 & 0.12 \\
HD 107146 & 29 & 100 &12:19:06 & +16:32:54 & G2 V & 5859 & 0.04 \\
\end{tabular}
}
\tablecomments{\cite{Hillenbrand}}
\label{tab:starprops}
\end{table*}
\end{center}
The Submillimeter Array (SMA) is an 8-element radio interferometer located on the summit of Mauna Kea at an altitude of nearly 13,800 ft.\footnote{The SMA is a collaborative project of the Smithsonian Astrophysical Observatory (SAO) and the Academia Sinica Institute of Astronomy and Astrophysics (ASIAA).} Each antenna is 6m in diameter. Our observations utilized the Subcompact, Compact, and Extended configurations of the SMA to sample baselines between 9 and 226 m, and to achieve an angular resolution as fine  as 2.2'' at 230 GHz and 0.6'' at 345 GHz. The sources were observed for two to four nights between 2008 and 2012. The SMA observations of HD 107146 have been previously published by \citet{Hughes11}, and observations of HD 61005 have been previously published by \citet{Ricarte13}; more detail about those data are available in their respective publications.  

Solar system bodies with well-determined flux models were used to calibrate the absolute flux scale of the observations. Quasars close to the target region were selected as gain calibrators, so that frequent reference could be made to them throughout each track to monitor changes in the instrumental and atmospheric gains. For source HD 104860, recalibration was necessary halfway through one track, so the additional gain calibrator and derived flux are provided. Table~\ref{tab:obparams} presents basic parameters of the SMA observations for each night of observation, including the number of antennas, baseline lengths, 225 GHz opacity (a measure of the transparency of the atmosphere), the RMS noise in the naturally weighted map, the synthesized beam size, the solar system object used as the flux calibrator for each track, the quasar used as the gain calibrator, and the flux derived for the gain calibrator. For all observations, the correlator was configured to provide maximum sensitivity across the full 8\,GHz available bandwidth.  
\begin{center}
\begin{table*}
\caption{SMA Observations}
\resizebox{\textwidth}{!}{
\begin{tabular}{ccccccccccc}
\hline\hline\noalign{\smallskip}
\textbf{Source} & Date(s) & Ant. & Baselines& $\tau_\mathrm{225\,GHz}$ & LO Freq& RMS Noise& Synth.& Flux & Gain & Derived \\ 
& & & (m) & & (GHz) & (mJy/beam) & Beam ('') &Cal & Cal&Flux (Jy) \\ 
\noalign{\smallskip}\hline\noalign{\smallskip}
HD 377$^\text{a}$ & 2012 Jun 25 & 7 & $16-77$ & $0.03-0.06$ & 340.794 & 0.64 & $2.16\times 1.98$ & Uranus & J0006-063& 0.66\\
       & 2012 Sep 04 & 7 & $44-226$ & $0.03-0.07$ & 340.773 & 0.64 & $0.76\times 0.55$& Neptune/Uranus & J0010+109 & 1.1\\ 
HD 8907$^\text{a}$ & 2012 Jun 29 & 7 & $16-77$ & $0.04-0.05$ & 225.477 & 0.30 & $3.61\times 3.18$ & Neptune &J0113+498& 0.43\\ 
	& 2012 Jul 24 &   &	& & & & $8.19\times7.60$& & & \\
	& 2012 Jul 25 &   &	& & & & $8.17\times7.52$& & & \\
	& 2012 Aug 14 &   &	& & & & $7.92\times7.68$& & & \\
	& 2012 Aug 16 &   &	& & & & $8.76\times7.22$& & & \\
HD 61005$^\text{b}$& 2008 Dec 16 & 7 & $16 -68$ & $0.10-0.15$ & 225.499 & 0.69 & $2.74 \times2.17$ & Uranus/Titan & J0747-331& 0.90\\
        & 2009 Dec 29 & 8 & $16 -77$ & 0.06 & 225.169 & 0.6 & $5.7 \times2.9$ & Uranus/Titan & J0747-331& 0.80\\
        & 2010 Apr 13 & 8 & $16 -69$ & 0.04 & 225.169 & 0.8 & $6.2 \times3.0$ & Titan & J0747-331& 1.02\\
        & 2012 Jan 29 & 7 & $50-226$ & $0.03-0.1$ & 225.482 & 0.69 & $2.74\times 2.17$ & Uranus/Callisto & J0747-331& 0.95 \\ 
HD 104860$^\text{a}$ &2012 Jan 12 & 7 & $9-45$& $0.05-0.1$ & 225.472 & 0.12 & $3.19\times 2.83$ & Titan & J1153+495; J1048+717& 0.994,0.70 \\ 
HD 107146$^\text{c}$ & 2009 Jan 06 & 8 & $16-69$ & 0.07 & 340.783 & 0.97 & $\cdots$& Titan & 3C274& 1.3 \\
          & 2009 Jan 21 & 8 & $9.5-69$ & $0.03-0.06$ & 340.783 & 0.97 & $\cdots$& Titan & 3C274& 1.2 \\
          & 2009 Feb 01 & 8 & $9.5-69$ & 0.06 & 340.783 & 0.97 & $\cdots$& Ceres & 3C274& 1.1 \\ 
          & 2009 May 02 & 7 & $25-139$ & 0.05 & 340.737 & 0.97 & $3.12 \times 2.52$ & Titan & 3C274&1.1
\end{tabular}
}
\tablecomments{$^\text{a}$This work; $^\text{b}$\cite{Ricarte13}; $^\text{c}$\cite{Hughes11}}
\label{tab:obparams}
\end{table*}
\end{center}
The dust continuum emission toward HD 104860 was observed at a wavelength of 1.3\,mm by the $15-$antenna CARMA array over the course of three nights. Mars and MWC349 were used to calibrate the flux, while the time-variable atmospheric and instrumental gains were calibrated with the quasar J0958+655. The weather was very good for the first two nights of observation with sky RMS (a measure of phase stability) values close to 80 $\mu$m and $\tau_{225\,\text{GHz}}\sim0.2$. The third night suffered from lower quality weather with $\tau_{225\,\text{GHz}}$ steadily increasing from 0.1 to 0.3 and sky RMS $\sim200\,\mu$m, although the early parts of the night were usable. The observation utilized the full 8\,GHz bandwidth of the CARMA correlator to maximize continuum sensitivity. 
\begin{center}
\begin{table*}
\caption{CARMA Observations}
\resizebox{\textwidth}{!}{
\begin{tabular}{cccccccccc}
\hline\hline\noalign{\smallskip}\noalign{\smallskip}
Source & Date(s) & Antennas & $\tau_\mathrm{230\,GHz}$ & LO Freq& RMS Noise& Synth.& Flux & Gain & Derived \\ 
	   &         &          &        & (GHz)  & mJy/beam   & Beam ('') & Cal & Cal &Flux (Jy)\\
\noalign{\smallskip}\hline\noalign{\smallskip}\noalign{\smallskip}
HD 104860	& 12 April 10	& 15 & 0.183 & 227.5343& 0.43& $2.83\times2.51$& Mars & 0958+655 & 0.53 \\ 
			& 12 May 8		& 15 & 0.159 & 227.5329& 0.49 & $\cdots$& Mars & 0958+655 & 0.80 \\ 
			& 12 June 5 	& 15 & 0.185 & 227.5343& 0.64 &$\cdots$ & MWC349$^1$ & 0958+655 & 0.88 \\ 
\end{tabular}
}
\tablecomments{$^1$A flux of 1.72 Jy was assumed.}
\end{table*}
\end{center}

The 1.3\,mm wavelength ALMA continuum data on HD 107146 originate from Cycle 0 observations by \citet{Ricci15}. The data collection and reduction processes are fully described in their paper. The main observational parameters for these observations are listed in Table~\ref{tab:aobparams}. For this work, visibility weights were estimated by calculating the variance of the real and imaginary parts of the visibilities across all channels in the data set, separately for each polarization and spectral window.  

\begin{center}
\begin{table*}
\caption{ALMA Observations$^\text{a}$}
\resizebox{\textwidth}{!}{
\begin{tabular}{cccccccccc}
\hline\hline\noalign{\smallskip}
Source & UT Date & Antennas & Baselines& pwv (mm) & LO Freq&  Flux & Bandpass& Gain \\ & & & (m) & & (GHz) & Cal & Cal & Cal \\
\noalign{\smallskip}\hline\noalign{\smallskip}\noalign{\smallskip}
HD 107146 & 11 Jan 12 & 17 & $19-269$& 2.29& 239.53&  Mars & 3C273& J1224+213 \\ 
	  & 27 Jan 12 & 16 & $19-269$& 3.02& 239.55&  Mars & 3C273& J1224+213 \\
	  & 27 Jan 12 & 16 & $19-269$& 2.86& 239.55&  Mars & 3C273& J1224+213 \\
	  & 16 Dec 12 & 23 & $15-382$& 1.13& 239.53&  Titan & 3C273& J1224+213  \\
	  & 01 Jan 13 & 24 & $15-402$& 2.82& 239.53&  Titan & 3C273& J1224+213  \\
\end{tabular}
}
\tablecomments{$^\text{a}$\cite{Ricci15}}
\label{tab:aobparams}
\end{table*}
\end{center}
\section{Analysis and Results}
\label{sec:analysis}

Spatially resolved observations of debris disks reveal the radial and azimuthal distribution of large dust grains. The spectral energy distribution allows us to constrain the temperature of the dust grains.  The combination of the two observables (SED + visibilities) is powerful because the spatial and thermal information, in conjunction with an assumption of standard astrosilicate opacities, allows us to infer basic properties of the size distribution of dust grains in the disk.  

In this section, we describe the theoretical framework we use to create a simple, but computationally efficient method of modeling the SED and visibilities (Section~\ref{sec:vismodel}).  The general modeling approach closely follows that described in \cite{Ricarte13}. The SED modeling utilizes a modified blackbody approach that will be described in detail below; similar approaches have been adopted by other authors to fit millimeter data and broad-band SEDs, including \citet{Williams04}, \citet{Hughes11}, and \citet{Booth}. While we do not implement a true grain size distribution, we approximate the effects of a grain size distribution through a combination of two parameters: the characteristic grain size $a$, analogous to the minimum grain size, and the long-wavelength slope of the grain absorption/emission efficiency $\beta$, which can be related to the power-law slope of the grain size distribution.  This approximation, which reduces the computational complexity of the code, decreases the run time per model by more than an order of magnitude, thereby allowing us to perform a robust statistical analysis of the data using an affine-invariant MCMC sampler (described in Section~\ref{sec:error}).

\subsection{SED and Visibility Modeling}
\label{sec:vismodel}
The SEDs are modeled with three components: 1) a Kurucz-Lejeune model photosphere, 2) a cold, spatially resolved outer debris disk, and when necessary, 3) a warm, compact dust belt. The Kurucz-Lejeune photosphere is not a free parameter in our model. Including the warm inner asteroid belt is necessary to reproduce the mid-IR fluxes in the SEDs for all the disks except HD 8907, since the contribution from the combination of the disk and star is insufficient to reproduce the overall flux in the mid-infrared. We used the Akaike information criterion (AIC) to place a $3\sigma$ upper limit on the warm belt mass of HD 8907 (see Table~\ref{tab:totparams}). The relative likelihood of a model with a second component is given by $\exp[(\text{AIC}_\text{1belt}-\text{AIC}_\text{2belts})/2]$, where $\text{AIC} = \chi^2 + 2k$, and $k$ is the number of parameters. The interpretation of two-temperature SEDs is discussed elsewhere in the literature in great detail \citep[see, e.g.,][]{Kennedy14}, and in some cases the presence of an inner asteroid belt may be degenerate with an additional population of small (and therefore hot) grains significantly smaller than the blowout size located in the outer belt, but for simplicity we assume that excess short-wavelength flux, when required, arises from spatially disparate populations of grains.  

We calculated the centroid of the disk using an elliptical Gaussian (or, for HD 107146, an elliptical ring) fit to the continuum visibilities with the MIRIAD\footnote{see \url{http://www.cfa.harvard.edu/sma/miriad/}} task {\tt uvfit}. All of the offset positions were consistent with the expected position of the star to within the uncertainties, taking into account the measured proper motion. We obtained initial flux estimates from these \texttt{uvfit} results as well, listed in Table~\ref{tab:diskprop}.  The position angle (PA) and inclination ($i$) used in the modeling process of HD 61005 and HD 107146 were adopted from scattered light observations of the disks (see \citealt{Buenzli10} and \citealt{Ardila04}), which are more precise than we were able to derive from the millimeter data alone. For HD 104860, the PA and $i$ were taken from \citealt{Morales13}; these values are consistent with (but slightly more precise than) the geometry derived from the major and minor axis lengths and position angle value calculated with \texttt{uvfit}. For HD 377 and HD 8907, the PA and $i$ were not well constrained by the \texttt{uvfit} results, so a grid search fit of PA and $i$ was used to maximize the visibility amplitudes; the results were consistent with the \texttt{uvfit} results.  These geometric parameters are highly uncertain; it is clear from the images and visibilities that these disks are marginally resolved by the interferometric data. Table~\ref{tab:diskprop} summarizes the measured fluxes and geometries for the five disks in the sample.
\begin{center}
\begin{table*}
\caption{Disk Properties}
\resizebox{\textwidth}{!}{
\begin{tabular}{lccccc}
\hline\hline\noalign{\smallskip}\noalign{\smallskip}
 Star      & Flux (mJy)  & Frequency (GHz)  & Position angle ($^\circ$) & Inclination ($^\circ$) & $R_\text{in}/R_\text{BB}$ \\ 
 \noalign{\smallskip}\hline\noalign{\smallskip}\noalign{\smallskip}
 HD 377    &$3.5\pm 1$   &345	&$30^a$		& $50^a$ & 1.6\\
 HD 8907   &$1.3 \pm 0.4$&230	&$55^a$		& $65^a$ & 1.7\\
 HD 61005  &$8.0\pm 0.8$ &230	&$70.3\pm1^b$&$84.3\pm 1^b$ & 6 \\
 HD 104860 &  $5.5\pm2$  &230	&$1 \pm 7^c$	& $52 \pm 6^c$ & 1.8\\ 
 HD 107146 & $70\pm20$   &345	&$58\pm5^d$		& $25\pm5^d$ & 3.8 \\
\end{tabular}
}
\tablecomments{$^a$These values were determined with a grid search. $^b$The position angle and inclination of HD 61005 \citep{Buenzli10}. $^c$The position angle and inclination of HD 104860 at 100$\,\mu$m \citep{Morales13}. $^d$The position angle and inclination of HD 107146 \citep{Ardila04}. For the last column, we report $R_\text{in}/R_\text{BB}$, where $R_\text{in}$ is determined through our modeling and fitting analysis, and $R_\text{BB}$ is determined by assuming the disk is radiating like blackbody and in equilibrium with its host star (see Section~\ref{subsec:rin} for details).}
\label{tab:diskprop}
\end{table*}
\end{center}
We calculate the temperature of the grains by assuming that they are in radiative equilibrium with the central star and that the disk is optically thin. The grains absorb and emit like graybodies with an absorption/emission efficiency $Q(a,\lambda)$ that is composition- and wavelength-dependent.

The efficiency is a function of $\lambda$ and the grain size $a$:
\begin{equation}
Q_{a,\lambda}=\frac{4}{3}\cdot\kappa_{tot}\cdot\rho\cdot a (1-\omega(\lambda))
\end{equation}
where $\kappa_{tot}(a,\lambda)$ is the grain opacity in $\cm^2\,\g^{-1}$, $\omega(a,\lambda)$ is the albedo or reflection coefficient, and $\rho$ is the grain density in $\g\,\cm^{-3}$. 

The grain opacity, $\kappa_{tot}(a,\lambda)$ can be calculated for spherical grains using predictions from Mie theory and geometric optics \citep{Draine06}. We assume that the grains are astronomical silicates (astrosilicates) with a bulk density of $2.7\, \g\,\cm^{-3}$ \citep{DraineLee84}. The energy per unit time absorbed by the grains, $\Gamma_\text{in}$ is
\begin{equation}
\Gamma_\text{in}=\pi a^2 \int_0^\infty Q(a,\lambda) \,F_\lambda(r,\lambda)\, d\lambda.
\end{equation}
where $F_\lambda$ is the flux density from a tabulated Kurucz model photosphere \citep{Lejeune97} drawn from the FEPS Legacy Survey archive\footnote{ \url{http://irsa.ipac.caltech.edu/data/SPITZER/FEPS/links.html}}.  The power emitted by the grains is the product of the Planck function, $B_\lambda(\lambda,T)$, and the emission efficiency, which equals the absorption efficiency: 
\begin{equation}
\Gamma_\text{out}=\pi a^2\cdot4\pi \int_0^\infty Q(a,\lambda) \, B_\lambda(T,\lambda)\, d\lambda. 
\end{equation} 
By setting $\Gamma_\text{in} = \Gamma_\text{out}$, we solve for the equilibrium temperature of the grains as a function of their size and distance from the central star.

We model each disk with a set of six parameters: an inner disk radius ($R_\text{in}$), a characteristic grain size ($a$), a disk mass ($M_D$), a warm inner belt mass ($M_B$), a grain emissivity parameter ($\beta$), and a disk width ($\Delta R$). $R_\text{in}$ affects the equilibrium temperature of the dust grains and is constrained by the disk visibilities. The characteristic grain size, $a$, determines the temperature of the grains. A smaller $a$ will shift the peak of the SED to shorter wavelengths as grains get hotter. $M_D$ is essentially a vertical scaling factor for the flux of the cold outer disk, which typically peaks at wavelengths of tens to hundreds of microns, while $M_B$ serves as a vertical scaling factor for the flux of the warm inner belt, which contributes most of its flux at shorter wavelengths. Given the limited information in the mid-infrared regime--primarily the fact that the inner belt has not been spatially resolved--the temperature and mass of the warm belt are highly degenerate, so we fix the temperature at 100 K and vary only the mass. As discussed in Section 4.2.1 in \cite{Ricarte13}, varying the temperature of the warm inner belt produces a noticeable change only in $M_B$. \cite{Ricarte13} demonstrate that the analysis of the cold belt is independent of the assumed belt temperature. The emissivity parameter $\beta$ determines the slope of the long-wavelength tail and can be related to the slope of the grain size distribution. $\Delta R$ describes the width of the outer belt and is also constrained primarily by the visibilities.  

We parameterize the surface mass density as $\Sigma(r)=\Sigma_{100}\cdot(r/100\,\AU)^{-p}$, where $\Sigma_{100}$ is the surface density in $\g\,\cm^{-2}$ at a distance of 100 AU from the central star and $p$ is the power law index that describes the radial dependence of surface density. There is a well-known degeneracy between $p$ and the outer radius (see section 4.2.2 of \citealt{Ricarte13}), but its effects are not significant in the case of a disk with spatially unresolved width. Since the radial width of disks in our sample is typically not spatially resolved, and therefore $p$ is not constrained, we fix $p$ at a value of 0 for all disks except HD 107146. \cite{Ricci15} show that the structure of HD 107146 warrants using $p$ as a free parameter. The surface number density of the grains, $N(r)$, is related to the surface mass density as $\Sigma(r) = N(r)m_g$, with $m_g = 2.7 \,\g\,\cm^{-3}\cdot 4\pi a^3/3$. 

When calculating the output SED, we approximate the grain emission efficiency following \cite{Williams04}: $Q(\lambda) = 1-\exp(-(\lambda_0/\lambda)^\beta)$, where $\lambda_0 = 2\pi a$ is the critical wavelength and $\beta$ is the opacity spectral index, instead of using the aforementioned tabulated astrosilicate opacities (which are used for the equilibrium temperature calculation only). This analytical parameterization is extremely computationally efficient and has the desired asymptotic properties for a grain size distribution with characteristic radius $a$, namely that $Q(\lambda)\approx(\lambda/\lambda_0)^{-\beta}$ for $\lambda \gg \lambda_0$ and $Q(\lambda)\approx1$ for $\lambda \ll\lambda_0$. While this parameterization of $Q$ preserves the asymptotic behavior, it smooths over features in the grain opacities used in the initial temperature calculations, so the code is not entirely internally self-consistent. The tradeoff is that the hybrid approximation to the grain size distribution allows our code to be efficient enough to run a thorough MCMC uncertainty analysis with a week of computing time on our local machines. The flux density at each wavelength is then, 
\begin{equation}
F_\lambda=\frac{\pi a^2 Q(\lambda)}{d^2}\int_{R_\textrm{in}}^{R_\textrm{in} + \Delta R} 2\pi r \,B_\lambda(T_r)\, N(r)\, dr 
\label{eq:pixflux}
\end{equation}
where $a$ is the characteristic grain size, $d$ is the distance to the star, $r$ is the distance of the grain from the star, and $N(r)$ is the surface number density of the grains.

To analyze the visibilities, we generate a high-resolution model image for comparison and calculate flux as a function of position, as described by equation~\ref{eq:pixflux}. The pixel size is set to be 1\% of the spatial scale sampled by the longest baseline.  We then sample the model image at the same spatial frequencies as the data, using the MIRIAD command \texttt{uvmodel}.  These model visibilities are compared with the observed visibilities in the visibility domain, using the appropriate statistical weights. 

\subsection{Error Analysis}
\label{sec:error}
We perform two separate chi-squared calculations, one comparing the model disk SED to photometric data from the literature (see table~\ref{tab:phot}) and a second comparing the disk visibilities to the model image of the disk. We add the chi-squared values for the SED and visibilities and use the total $\chi^2$ as the statistic for goodness-of-fit: $\chi^2 = \chi_\text{SED}^2+\chi^2_\text{VIS}$. As discussed in \cite{Andrews09}, the high quality of the low number of SED points balances the large numbers of visibilities, so that neither the SED nor the visibilities dominate the final fit. The uncertainties in the parameters of the fit can be determined through the use of a probabilistic sampling algorithm. 
\begin{center}
\begin{table*}
\caption{Broad-band photometry for all sources (flux densities in units of mJy)}
\resizebox{\textwidth}{!}{
\begin{tabular}{crclrclrclrclrclc}
\hline\hline\noalign{\smallskip}
$\lambda\,(\mu\m)$ & \multicolumn{3}{c}{HD 377} &\multicolumn{3}{c}{HD 8907}& \multicolumn{3}{c}{HD 61005} & \multicolumn{3}{c}{HD 104860} &\multicolumn{3}{c}{HD 107146} & Ref.\\
\noalign{\smallskip}\hline\noalign{\smallskip}
$1.24 $ & $4260$&$\pm$&$80.0$ & $8380$&$\pm$&$180$ & $2730$&$\pm$&$70$ & $2940$&$\pm$&$50$ & $7100$&$\pm$&$150$&a\\
$1.65 $ & $3640$&$\pm$&$70.$ & $6720$&$\pm$&$110$ & $2450$&$\pm$&$100$ & $2450$&$\pm$&$40$ & $5980$&$\pm$&$110$&a\\
$2.16 $ & $2380$&$\pm$&$50.0$ & $4510$&$\pm$&$70$ & $1740$&$\pm$&$40$ & $1670$&$\pm$&$30$ & $4040$&$\pm$&$60$ &a\\
$3.35 $ & $1160$&$\pm$&$50.0$ & $2100$&$\pm$&$120$ & $819.0$&$\pm$&$32.0$& $806$&$\pm$&$32$ & $1870$&$\pm$&$110$ &b\\
$3.6^\text{a}$ & $1029.1 $&$\pm$&$ 22.1$ & $1918.2$&$\pm$&$41.2 $ & $753.5$&$\pm$&$16.2 $ & $ 724.8$&$\pm$&$15.6 $ &$1711.3$&$\pm$&$36.7 $& c\\
$4.5^\text{a}$ & $ 648.6$&$\pm$&$14.9$ & $1223.7$&$\pm$&$28.1 $ & $472.3$&$\pm$&$10.8 $ & $455.3$&$\pm$&$10.5 $ &$1074.8$&$\pm$&$24.7 $& c \\
$4.60 $ & $651.0$&$\pm$&$13.0$ & $1380$&$\pm$&$30$ & $453.0$&$\pm$&$9.0$ & $442$&$\pm$&$9$ & $1230$&$\pm$&$30$ &b\\
$8.0^\text{a}$ & $234.7$&$\pm$&$5.0$& $427.3$&$\pm$&$9.1$ & $169.2$&$\pm$&$3.6 $ & $162.5$&$\pm$&$3.5 $ & $384.4$&$\pm$&$8.2 $ & c\\ 
$11.6 $ & $105.0$&$\pm$&$1.0$ & $193$&$\pm$&$2.0$ & $78.3$&$\pm$&$1.1$ & $74.3$&$\pm$&$1.0$ & $176$&$\pm$&$2$ &b\\
$13 ^\text{a}$ & $81.6$&$\pm$&$5.0 $ & $ 154.1$&$\pm$&$9.4 $& $62.3$&$\pm$&$3.8 $ & $57.3$&$\pm$&$3.5$ & $138.9$&$\pm$&$8.5 $ & c\\
$22.1 $ & $41.7$&$\pm$&$1.7$ & $61.5$&$\pm$&$1.6$ & $44.4$&$\pm$&$1.6$ & $23.1$&$\pm$&$1.1$ & $69.6$&$\pm$&$1.9$ &b\\
$ 24^\text{a}$ & $36.6$&$\pm$&$1.5 $ & $51.3$&$\pm$&$2.1 $ & $41.5$&$\pm$&$1.7 $ & $ 19.9$&$\pm$&$0.8 $ & $59.8$&$\pm$&$2.5 $ & c\\
$33^\text{a}$ & $37.8$&$\pm$&$2.7 $ & $41.8$&$\pm$&$3.5 $ & $110.0$&$\pm$&$6.7 $ & $17.8$&$\pm$&$1.8$ & $86.7$&$\pm$&$5.7$ & c\\
$70^\text{a}$ & $162.0$&$\pm$&$16.9$& $247.4$&$\pm$&$19.7 $ & $628.7$&$\pm$&$45.4 $ & $ 183.1$&$\pm$&$14.8 $ & $ 669.1$&$\pm$&$47.8$ & c\\
$100^\text{b}$ & & $ \cdots$ & & &$ \cdots$& & &$ \cdots$& & $277.0$&$\pm$&$3.5$& &$ \cdots$& &d \\
$160^\text{a}$ & $ 187.5$&$\pm$&$50.4 $& $243.8$&$\pm$&$42.3$& $502.6$&$\pm$&$160.1$& $202.7$&$\pm$&$27.0$& &$ \cdots$ & & c\\
$160^\text{b}$ & & $ \cdots$ & & &$ \cdots$& & &$ \cdots$& & $243.4$&$\pm$&$5.2$& &$ \cdots$& & d\\
$350$ & &$\cdots$& & &$\cdots$& & $95$&$\pm$&23 & $50.1$&$\pm$&$14$& 319&$ \pm$& 64 &  e\\
$450$ & &$\cdots$& & $22$&$\pm$&$13$& &$\cdots$& & $47$&$\pm$&$18$& 130&$ \pm$& 40 & f\\
$850$ & &$\cdots$& & $4.8$&$\pm$&$1.3$& &$\cdots$& & $6.8$&$\pm$&$1.4$& 20&$ \pm$ &4 &f \\
$880^\text{c}$& 3.5&$\pm$&1.4& &$\cdots$& & &$\cdots$& &  &$\cdots$& & 36&$\pm$&1 & g\\
$1200$ & 4.0 &$\pm$&1.2 & $3.2$&$\pm$&$1.0$& &$\cdots$& & $4.4$&$\pm$&$1.3$& &$ \cdots$ & & e\\
$1300^\text{c}$ & &$\cdots$& & $1.3$&$\pm$&$0.4$& 7.98&$\pm$&0.8 & 5.5 &$\pm$&1.8& &$ \cdots$ & & g \\
$1300^\text{c}$ & &$\cdots$& & &$\cdots$&& &$\cdots$& & &$\cdots$&& 12.5&$ \pm$ &1.3 &h \\
\end{tabular}
}
\tablecomments{$^\text{a}$ 2MASS, \citep{Kharchenko09}; $^\text{b}$WISE, \citep{Cutri12}; $^\text{c}$FEPS \citep{Hillenbrand}: The uncertainties include both internal and calibration terms. $^\text{d}$Herschel photometry \citep{Morales13}. $^\text{e}$CSO and IRAM \citep{Rocca09}: The uncertainties include additional 20\% calibration uncertainties on the 350$\,\mu$m fluxes and 16\% calibration errors on the 1200$\,\mu$m fluxes. $^\text{f}$JCMT/SCUBA \citep{NajWill}. The uncertainties include additional 30\% calibration uncertainties on the 450$\,\mu$m fluxes and 10\% calibration errors on the 850$\,\mu$m fluxes. $^\text{g}$From this work. $^\text{h}$ALMA \citep{Ricci15}, The uncertainty reflects a 10\% systematic flux uncertainty.}
\label{tab:phot}
\end{table*}
\end{center}
The Markov Chain Monte Carlo (MCMC) technique is a random sampling algorithm \citep{Recipes} that provides a powerful method of determining the uncertainties on the model parameters, taking into account both the uncertainties on individual data points and degeneracies between parameters in the model. We utilize the affine-invariant MCMC fitting technique described in \citet{Goodman}. 

Affine invariant sampling efficiently eores degenerate parameter spaces due to its lack of bias in treating distributions that are highly anisotropic. Assigning a dimension to each model parameter, the resulting $N$-dimensional space is initialized with a uniformly distributed set of walkers, or vectors in the space that contain parameter values. In the \cite{Goodman} \textit{stretch move}, the walkers (sets of model parameters) explore the space by moving along lines containing other walkers. The decision for a walker to explore the space depends on whether the likelihood function will be maximized by this move.  We implement a sampler that utilizes only stretch moves, similar to the affine-invariant MCMC sampler, \texttt{emcee}, written in Python \citet{ForemanMackey}. Online documentation and a full description of the \texttt{emcee} likelihood function are available at \url{http://dan.iel.fm/emcee/current/}.  We assume uniform priors for all variables.

In the initialization of the walkers, trial states for $a$, $M_D$, and $M_B$ are generated in logarithmic space, while states for $R_\text{in}$ and $\beta$ (and $p$ for HD 107146), are generated in linear space. Please see Section 4.2 of Ricarte et al. (2013) for a discussion of  the parameter degeneracies. We run 50 walkers through 5000 trials and discard trials for which the $\chi^2$ values have not yet converged (the ``burn-in" phase, typically a few hundred trials). The posterior probability density functions for each parameter, marginalized over all other parameters, are calculated from the ensemble of walkers across all trials excluding burn-in.

\subsection{Results}
\label{sec:results}

\begin{figure*}
\centering
\includegraphics[scale=0.8]{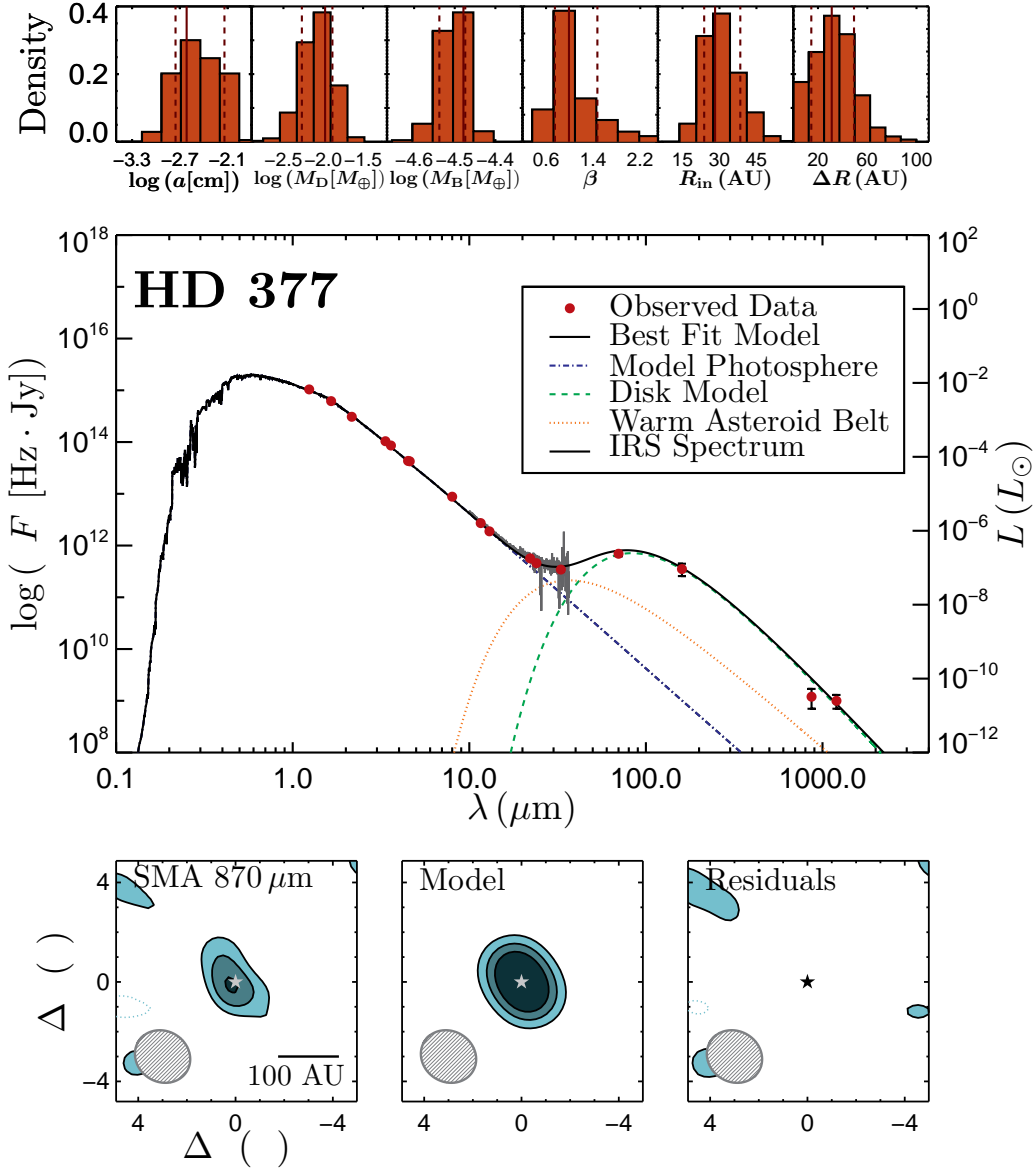}
\caption{Modeling results for HD 377.  Top: Marginalized posterior probability density functions (PDFs) for the model parameters as derived from the MCMC analysis. The dashed gray vertical lines mark the $\pm1\,\sigma$ range that encloses models within one standard deviation of the most probable value. The solid line marks the most probable value determined from the analysis. Middle: SED of the system. The total model SED is the sum of three components: a Kurucz-Lejeune model photosphere, a modified blackbody modeled to the debris belt, and a warm asteroid belt. The IRS spectra are not included in the modeling process, yet they provide visual check of the total model at mid-infrared wavelengths. Bottom: The interferometric image of the emission from the disk around each star. The image has been constructed using data with all-array configurations. Contours are drawn at $[2,3,4]\times0.7$ mJy beam$^{-1}$ (the rms noise). The axes have been set such that the (0,0) position corresponds to the expected position (corrected for proper motion), of the star, which is marked with a $\star$ symbol. A gray ellipse indicating the size of the synthesized beam is drawn in the lower left corner. The black bar in the lower right corner illustrates the linear scale.}
\label{fig:fullhd377}
\end{figure*}

\begin{figure*}

\begin{center}
\includegraphics[scale=0.95]{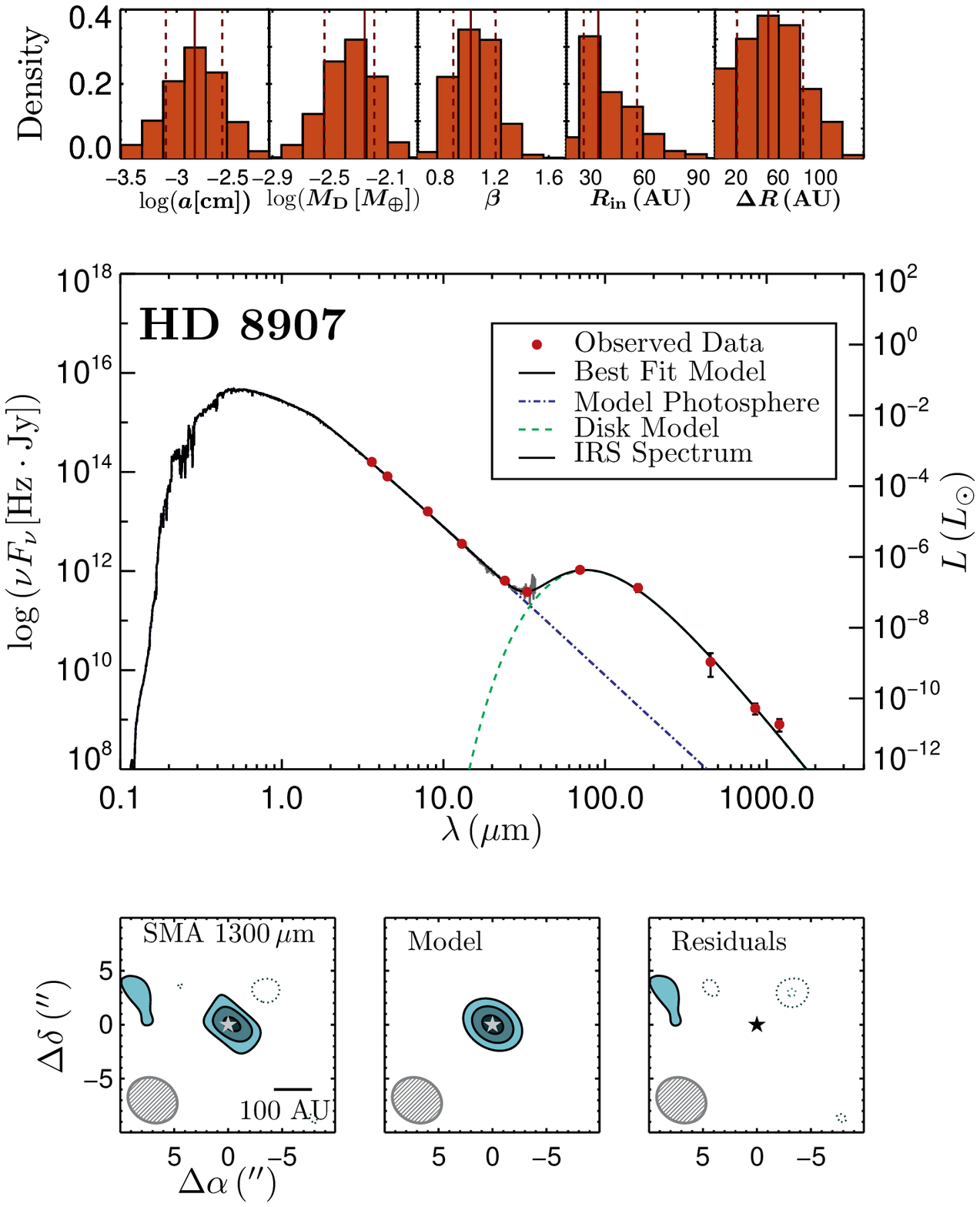}
\caption{Same as Figure~\ref{fig:fullhd377}, but for HD 8907. Contours are drawn at $[2,3,4]\times0.3$ mJy beam$^{-1}$.}
\label{fig:fullhd8907}
\end{center}
\end{figure*}

\begin{figure*}
\centering
\includegraphics[scale=0.9]{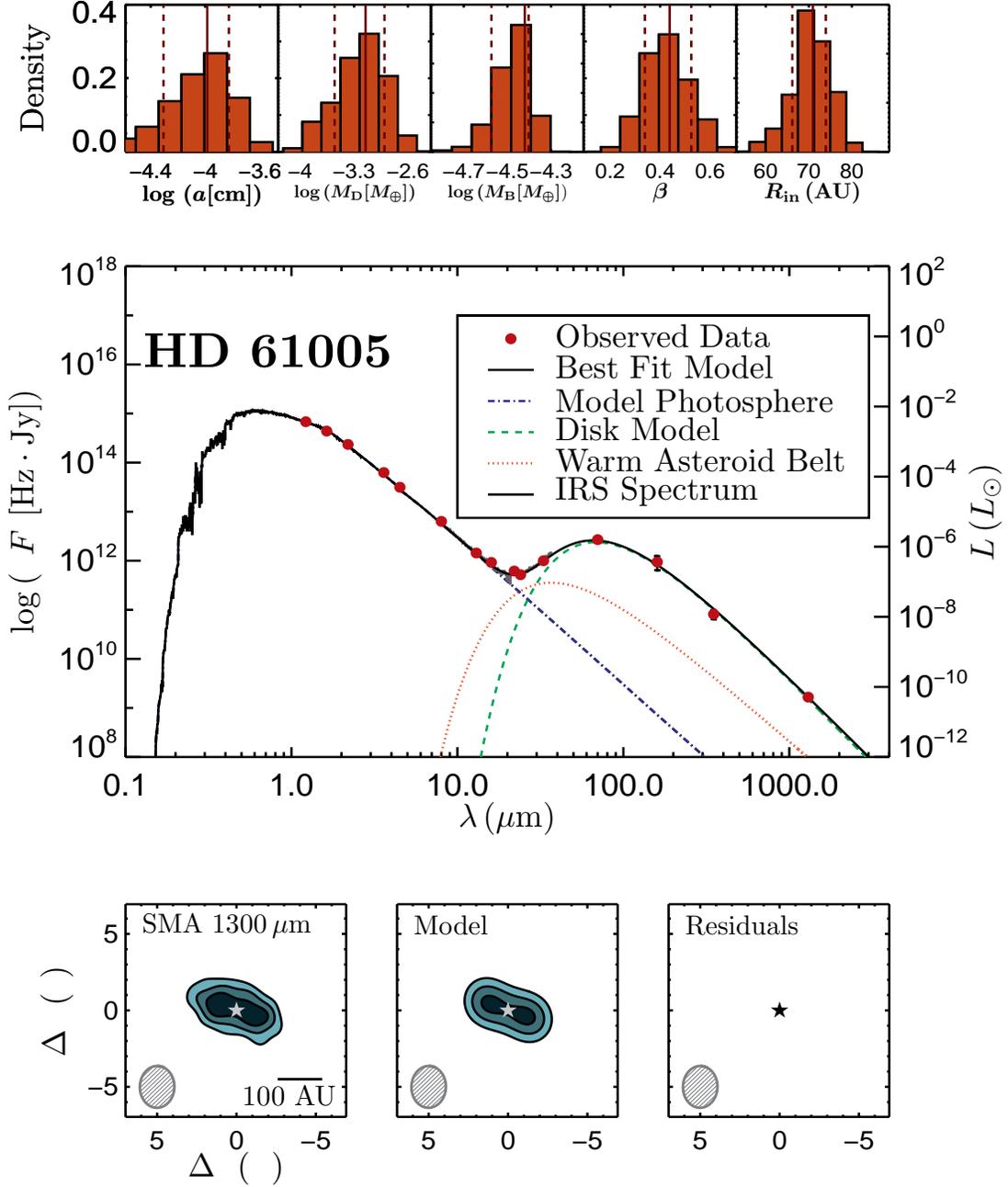}
\caption{Same as Figure~\ref{fig:fullhd377}, but for HD 61005. Contours are drawn at $[3,5,7]\times0.4$ mJy beam$^{-1}$.}
\label{fig:fullhd61005}
\end{figure*}

\begin{figure*}
\begin{center}
\includegraphics[scale=0.9]{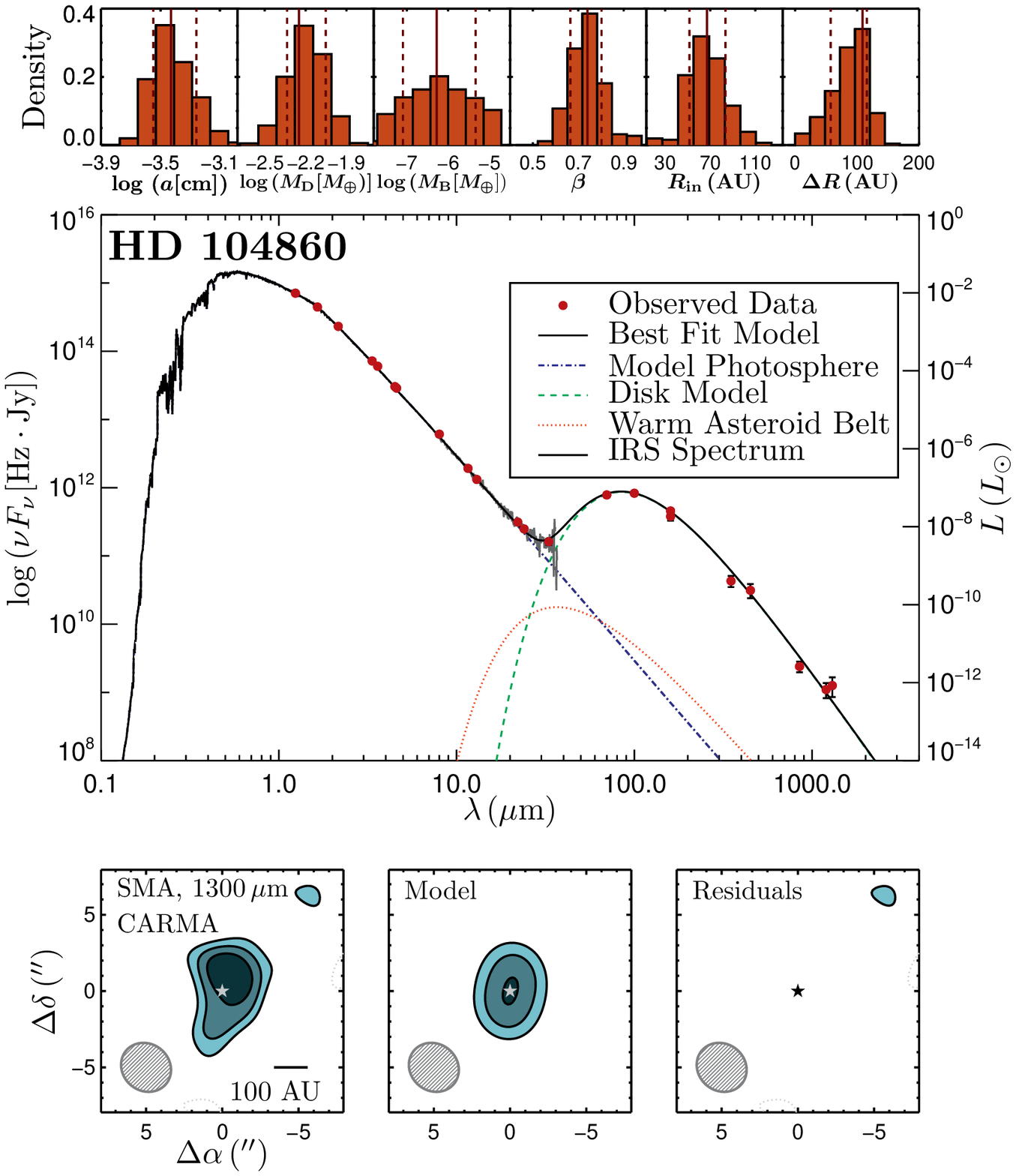}
\caption{Same as Figure~\ref{fig:fullhd377}, but for HD 104860. Contours are drawn at $[2,3,4]\times0.3$ mJy beam$^{-1}$.}
\label{fig:fullhd104860}
\end{center}
\end{figure*}

\begin{figure*}
\begin{center}
\includegraphics[scale=0.85]{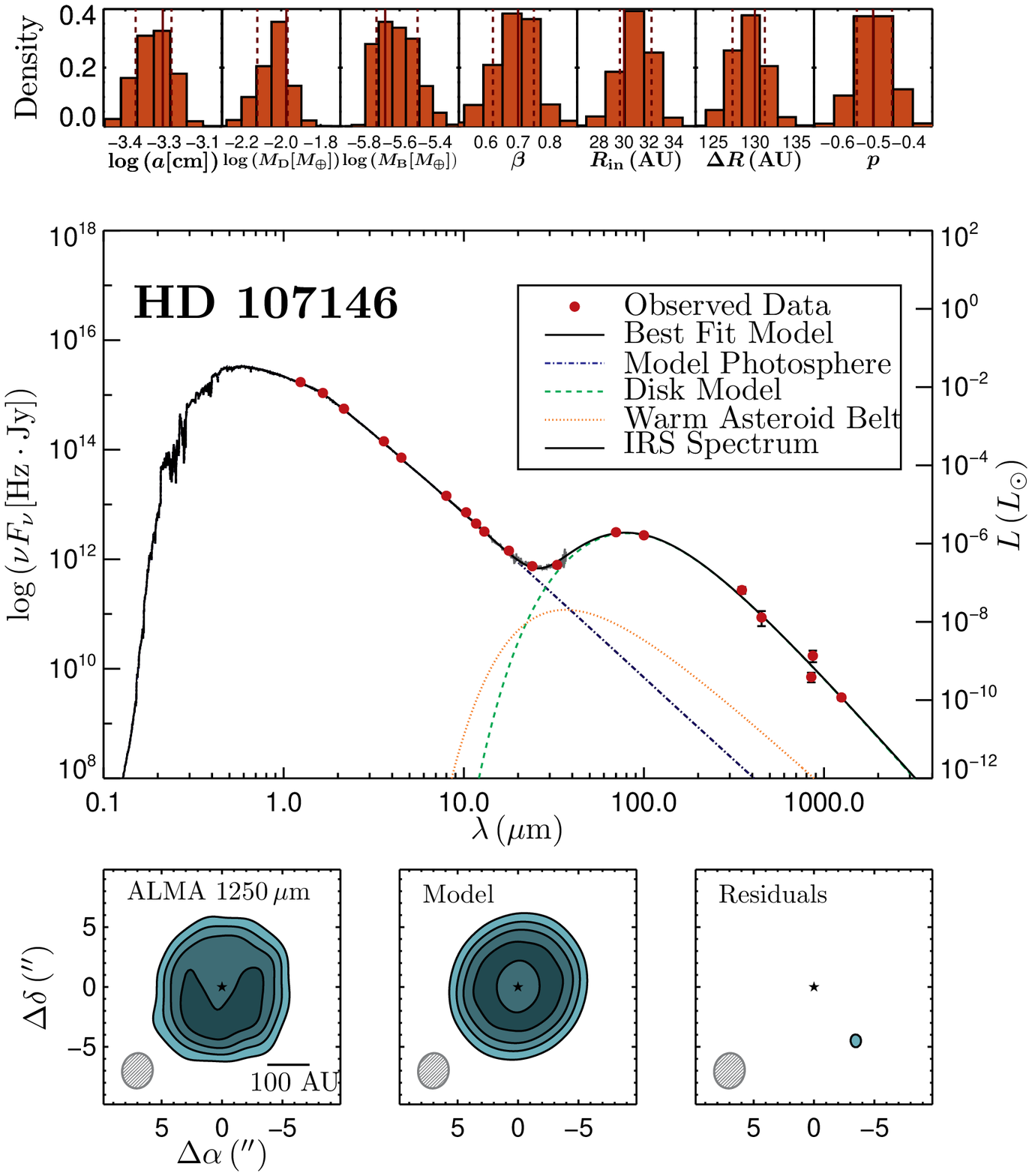}
\caption{Same as Figure~\ref{fig:fullhd377}, but for HD 107146. Due to the greater dynamic range, the contours are drawn at $[3,8,13,18]\times0.08$ mJy beam$^{-1}$. The additional parameter $p$ is the slope of the radial power law of the parameterized surface mass density (see Section~\ref{sec:vismodel}). We discuss the implications of this parameter in Section~\ref{sec:vismodel}.}
\label{fig:fullhd107146}
\end{center}
\end{figure*}

By simultaneously modeling the SED and the visibilities, we constrain basic properties for each disk in our sample. Figures \ref{fig:fullhd377}-\ref{fig:fullhd107146} show the posterior probability density functions and the best-fit SED and images for each source. The global best fit model is chosen from the MCMC walker position with the lowest $\chi^2$ value across the entire set of walkers and trials. To generate residual images, we subtract the model from the data in the visibility domain, and then image the residual visibilities using the same imaging parameters as for the data and model images.  The most probable value in the posterior PDF typically corresponds well with the best-fit value. The uncertainties on the best-fit parameters reported in Table~\ref{tab:totparams} represent the range of values in the PDF which encloses $1\sigma$ (68\%) of the models. The errors are not consistently symmetric about the best fit. 

\begin{table*}
\begin{center}
\caption{The Best-fit Parameters for All Disks}
\resizebox{\textwidth}{!}{
\begin{tabular}{lcccccccc| c c }
\hline\noalign{\smallskip}\noalign{\smallskip}\
Source    & $a(\mu\m)$ & $M_D(10^{-3}M_\oplus)$ & $M_B(10^{-5}M_\oplus)$ & $\beta$ & $R_\text{in}$(AU)& $R_\text{in,avg}$ & $\Delta R$ (AU) &$ p$ & Total $\chi^2$&  $\chi^2_\text{Red}$ \\ 
\noalign{\smallskip}\hline\noalign{\smallskip}
HD 377	& $19.8$	& $5.57$	& $3.26$ 	&$0.76$ 	& $31$ & 47 & $32$&-- & 269023.48 & 1.40 \\
		& $33.5^{+39}_{-16}$	& $8.38^{+4.64}_{-2.99}$& $3.13^{0.33}_{-0.36}$ &$0.96^{+0.51}_{-0.21}$ & $30.1^{+8.4}_{-6.3}$ & & $32.9^{+16}_{-18}$&-- \\
HD 8907   & $24.9$  & $4.93$  & $2.68\times10^{-5} (^\text{a})$ &$1.13$ & $28$ &$54$ &$52$&--& 357244.56 & 1.89 \\
			&$15.4^{+13}_{-7.6}$  & $4.49^{+2.05}_{-1.57}$  & -- &$1.05^{+0.16}_{-0.15}$ & $36.4^{+19}_{-10}$ & &$53.2^{+31}_{-32}$&-- \\
HD 61005  & $1.17$ & $0.854$ & $2.99$  & $0.46$ & $69.4$& 71 &--$^\text{b}$ & -- & 456165.08 & 1.85 \\
			& $1.03^{0.46}_{0.54}$ & $0.71^{+0.54}_{-0.43}$ & $4.26^{+0.17}_{1.3}$  & $0.43\pm 0.08$ & $70.9^{+3.0}_{-4.7}$& & --$^\text{b}$ & -- \\

HD 104860 & $4.27$ & $7.20$ & $1.68$ & $0.75$ & $57$ & 110 &$108$ & -- & 377760.86 & 2.38 \\
			& $4.00^{1.63}_{-1.12}$ & $6.87^{+2.42}_{-2.17}$ & $0.193^{+2.3}_{-0.019}$ & $0.74\pm 0.08$ & $63.3^{+24}_{-11}$ & & $87.9^{+24}_{-43}$ & --  \\
HD 107146 & $4.71$& $8.85$ & $0.197$  &$0.74$ & $29.4$& 94 &$129$ & $-0.57$ & 490862.26 & 1.18\\
			& $5.05^{+0.81}_{0.78}$& $9.51^{+1.27}_{-1.95}$ & $0.241^{+0.069}_{-0.049}$  &$0.75^{+0.02}_{-0.05}$ & $30.8^{+2.0}_{-1.7}$& & $129^{+2.1}_{-1.9}$ & $-0.50^{+0.08}_{-0.07}$\\
\hline
\end{tabular}
}
\label{tab:totparams}
\tablecomments{$^\text{a}$This belt mass is an upper limit based on a 3$\sigma$ significance to the best fit with one (cold) belt. $^\text{b}$To maintain consistency with \citet{Ricarte13}, the width of the belt was fixed at 5\% of the inner radius. The top row for each source gives the global $\chi^2$ minimum and the second row gives the median $\pm$ 1$\sigma$ uncertainty. $R_\text{in,avg}$ is defined as $R_\text{in} + \Delta R/2$.  We include it in the table for easier comparison with other models that report the midpoint as the radius of the ring.}
\end{center}
\end{table*}


\section{Discussion}
\label{sec:discussion}

\subsection{Grain Sizes}
The disks have characteristic grain sizes ranging from $\sim\!1\,\mu$m to $\sim\!30\,\mu$m. We can estimate a minimum grain size that should be observable in a debris disk by calculating the grain size for which radiation pressure balances the gravitational force exerted on the orbiting dust grain: 
\begin{equation}
a_\text{blow} = \frac{3\,L_\star}{16\pi G M_\star c\rho}
\end{equation}
where $L_\star$ is the stellar luminosity, $M_\star$ is the mass of the star, $c$ is the speed of light, and $\rho$ is the grain density (e.g., \citealt{Backman}). Grains smaller than this blowout size are efficiently removed from the disk on timescales much shorter than the age of the star. The average blowout sizes here are $\sim0.3\,\mu$m, roughly an order of magnitude smaller than the average characteristic grain size. All of the SEDs are therefore reproduced well with distributions that only include grains larger than the blowout grain size. HD 377 and HD 8907 have slightly larger characteristic grain sizes of $\sim\!10\,\mu$m, which may point toward grain growth, although these are the most poorly spatially resolved disks in the sample and the grain sizes in these systems are therefore particularly uncertain.

$\beta$ controls the long-wavelength slope of the SED and reflects the slope of the grain size distribution, or the number of small grains compared to large grains in the disk \citep{Wyatt08e}.  Typical $\beta$ values observed in protoplanetary disks are $\beta=2$ for interstellar medium grains, $0<\beta<1$ for pebbles of the order of 1 mm, and $\beta=0$ for large grains \citep{Beckwith}. The values of $0<\beta<1$ for most debris disks point toward grain growth.  To connect the measured values of $\beta$ to the slope of the grain size distribution, we use the relationship derived by \citet{Draine}: $\beta \approx (q - 3) \beta_s$, where $\beta_s$ is the dust opacity spectral index in the small particle limit, which \citet{Draine} find to be $\approx 1.8\pm0.2$, and $q$ is the slope of the grain size distribution $\text{d}n(a) \propto a^{-q} \text{d}a$.  Solving for $q$, we find that $q \approx (\beta/\beta_s) + 3$.  The median value of beta in our sample was $\sim\!0.9$, with typical uncertainties of $\pm\sim\!0.1$ depending on the sampling of the long-wavelength portion of the SED.  We therefore derive a typical $q$ value of $3.5\pm 0.5$.  This result is consistent with the $q$ values anticipated for most theoretical models of collisional cascades, including $q = 3.51$ \citep{dohnanyi69}, $q = 3.65$ \citep[for the fiducial model presented in][]{gaspar11}, as well as the range of values derived by \citet{pan12}.  The $\beta$ values in our sample are therefore typical of what we expect of the grain size distribution in a disk undergoing a collisional cascade.

\subsection{Inner Radii and Disk Widths}
\label{subsec:rin}
The inner radii of the disks in our sample range from 28\,AU to 69\,AU with typical uncertainties of $\pm10$ AU.  The median value was 42\,AU, and all but one of the disks (HD 61005) was consistent with that median value to within the uncertainties. 

Looking at a sample of 34 debris disks around stars of spectral types A through M resolved in various \textit{Herschel} programs (including HD 104860, Vega, AU Mic, and Fomalhaut), \cite{Pawellek14} modeled the SEDS and Herschel-resolved images of debris disks, focusing on the cold, outer component. They report radii ranging from 40\,AU to 290\,AU, with only weak correlation to luminosity of the host stars. Of the 21\% of disks with radii $<\!100$ AU, the median radius is 60 AU, while for the $79\%$ with radii $>\!100$ AU the median radius is 154 AU. Large inner radii can hint at the presence of fully formed planets and would reflect the semimajor axis distributions of those planets. However, these disks were resolved at much shorter wavelengths and may not trace the parent planetesimal belts as reliably as millimeter wavelengths.  In addition, the authors compare a modified blackbody fitting method to a full grain size distribution method and find that while there are some quantitative differences between the results, the trends in grain temperatures and disk radii are consistent across the two methods.  They suggest that a modified blackbody approach like the one used here is simple, transparent, and may be appropriate for stars with relatively sparse SEDs.  

The radius of the disk is commonly estimated using two distinct methods. $R_\text{BB}$ represents the radius estimated from the SED, assuming that grains are in blackbody equilibrium with the central star, while $R_\text{in}$ represents the inner radius determined by a simultaneous fit to the SED and visibilities, assuming modified blackbody grains.  The ratio of $R_\text{in}$ to $R_\text{BB}$ is typically greater than one since the small grains are hotter than the blackbody equilibrium temperature. \cite{Morales13} compare the disk sizes resolved by \textit{Herschel} relative to the expectation from blackbody emission and find that the ratio of resolved radii to blackbody equilibrium radii depends on the luminosity of the star. We find that the radii resolved in this work are consistent with typical values observed for solar-type stars. Radiation pressure will clear a disk of its smallest grains, so the greater the host star's luminosity, the closer the ratio $R_\text{in}/R_\text{BB}$ is to unity.  We find that a solar analogue disk tends to have a ratio $R_\text{in}/R_\text{BB} \sim2$ (see Table ~\ref{tab:diskprop}), consistent with the values determined for Solar-type stars by \citet{Morales13}. \cite{Pawellek14} also find that the ratio is slightly greater than one using a modified blackbody approach and $\sim\!2$ using a size distribution at infrared wavelengths.

From the mid-IR portions of the SEDs, we find that at least three disks in this sample require two-temperature belts, which we interpret as two radially distinct debris belts, similar to \citet{Chen06}, \citet{Chen09}, \citet{Su09}, and \citet{Morales11}. \cite{Chen14} find that the majority of sources with 13, 31, and $70\mu$m excesses require multiple components to simultaneously fit the observed range of excesses. The debris disk around HD 104860 has been fit with one component \citep{Pawellek14} and two components \citep{Morales13}.  We analyzed the SED and visibilities with one component as well as two, and find that a two-component disk best reproduces the thermal SED with at least a 3$\sigma$ significance using the Bayesian Information Criterion (BIC).  HD 8907 is the only disk in the sample that clearly requires only one component, yet we also place an upper limit on the warm belt mass (see Table ~\ref{tab:totparams}). The lack of excess warm dust implies that this disk has an inner radius of at least tens of AU.


HD 107146 is the only disk in this sample with an unambiguously resolved disk width. Recent ALMA results of HD 107146 demonstrate that in fact its disk surface density is likely quite complicated with a deficit of flux between the inner and outer radii \citep{Ricci15}. The other disks around solar-type (F and G types) have widths that are smaller than their inner radii $(<\!\!100\,$AU) and are not spatially resolved by the observations.  For comparison, the Solar System's Kuiper belt begins at a radial distance of 40\,AU from the Sun.  The ``classical" Kuiper belt truncates at a distance of 50\,AU and its ``scattered" component extends for hundreds of AU \citep[see review articles in][]{kuiper2009}.  

Significant differences in radial width at different wavelengths have now been observed for several different debris disk systems.  Two particularly striking examples are the edge-on disks around AU Mic and $\beta$ Pic, which exhibit narrow millimeter-wavelength rings embedded in significantly broader scattered light distributions \citep{Smith84,Kalas04,Krist05,Wilner11, Wilner12, Kalas13,Schneider14,Apai15}.  The location of the narrow millimeter ring corresponds with the radius of a break in the surface brightness power law observed in scattered light, which dovetails with theoretical predictions by \cite{Strubbe}.  Their analysis of the brightness distribution of AU Mic predicts the presence of a ``birth ring'' of large planetesimals at the radius of the break in the scattered light power law.  Because small grains are influenced by the effects of radiation pressure, they are blown to much larger radii, creating a ``halo" around the birth ring.  The millimeter grains, by contrast, are far less affected by stellar radiation pressure, and more faithfully trace the location of the parent planetesimal belt.

At the time of writing, only two of the disks in our sample have been resolved in scattered light, namely HD 61005 and HD 107146 \citep{Soummer14,Schneider14}. The disk around HD 107146 exhibits only a slight extension of scattered light beyond the extent of the millimeter grains \citep{Ertel11,Ricci15}.  The ratio of the width of the disk observed in scattered light, $\Delta R_\text{sl}$, to the width of the disk observed at millimeter wavelengths, $\Delta R_\text{mm}$, is approximately 1.2 (given the approximate inner radius of 60 AU and outer radius of 220 AU quoted for the scattered light extent of HD 107146; see Table 5 in \citealt{Schneider14}).  By contrast, both the AU Mic and $\beta$ Pic debris disks exhibit scattered light haloes that extend to several times the radius of their millimeter counterparts \citep{Wilner11,Wilner12}, although AU Mic's millimeter disk exhibits a low surface brightness component that extends in toward the central star with no detectable inner radius \citep{MacGregor13}.  From analyses of the HD 61005 disk (\citealt{Ricarte13,Buenzli10,Hines07}), the scattered light images can reveal external physical processes that shape the disk. For example, \citet{Ricarte13} show that the millimeter grains are more strongly confined to the parent planetesimal belt than the more spatially extended scattered light morphology, providing support for an ISM-related origin to the dramatic swept-back morphology of the disk.  

\subsection{Deviations from an Axisymmetric Model}
Non-axisymmetric features such as eccentric rings, warps, and spiral arms, have been observed in several debris disks, primarily at optical and near-IR wavelengths.  These features are generally thought to be caused by dynamical interactions between the disk and its (usually unseen) planets.  In at least two cases ($\beta$ Pic: \citealt{Smith84,Mouillet97,Lagrange10,Currie11,Lagrange12a}; and Fomalhaut: \citealt{Wyatt02f,Holland03,Quillen06,Chiang09,Janson12}), such features have pointed the way to the discovery of directly imaged planetary-mass companions orbiting the central star.  While other mechanisms have been proposed to explain some of these non-axisymmetric features (for example, gas pressure gradients may cause eccentric rings \citep{Lyra13}; stellar flybys may cause warping (\citealt{Zakamska04,Malmberg07,Malmberg09,Malmberg11,Marzari13}); and interactions with the ISM may cause large-scale asymmetries including swept-back structure (\citealt{Debes09, Maness10}), the success of direct imaging studies demonstrates that at least in some cases structure in debris disks does reveal the presence of planets far from their host star. 

A long-standing theoretical prediction holds that millimeter wavelengths, with their sensitivity to macroscopic particles that are insensitive to the effects of stellar radiation pressure, should be ideal for observing resonant clumpy structure generated by resonant interactions between a planet and nearby dusty debris (\citealt{Wyatt03, Wyatt06, Wyatt08}).  This prediction is consistent with the observation that Kuiper belt objects have been trapped in Neptune's major resonances, likely as a result of its past migration through the early solar system's planetesimal disk (\citealt{Chiang03, Hahn05}).   However, attempts to detect clumpy structure in debris disks using millimeter wavelength interferometry have a somewhat checkered history: early observations of clumps consistent with resonances in Vega's debris disk (\citealt{Koerner01, Wilner02}) were not confirmed using more sensitive observations (\citealt{Pietu11, Hughes12}), and apparent clumpy structure in the HD 107146 debris disk \citep{Corder09} was later demonstrated to be consistent with random noise in low-S/N images \citep{Hughes11}.  To date, the only dust continuum asymmetry observed in a debris disk observed using millimeter-wavelength interferometry is a relatively subtle brightness asymmetry between the two ansae of the edge-on $\beta$ Pictoris debris disk \citep{Dent14}.  All other interferometric observations of debris disks, including several sensitive studies with the ALMA observatory (\citealt{Boley12, MacGregor13}, \citealt{Ricci15}), have been well reproduced by a smooth, axisymmetric density distribution. It is also worth noting that there is no evidence for clumpy structure in the HR 8799 debris disk (\citealt{Patience11, Hughes11}, Booth et al. in prep.), despite the known presence of at least four giant planets orbiting just interior to the outer dust disk (\citealt{Marois08, Marois10}).  

Our results, which demonstrate that all five disks in our sample are similarly well reproduced by axisymmetric density distributions, are in line with these previous results. There is a localized 3-sigma peak in the ALMA HD 107146 residual map, although a single 3$\sigma$ peak is consistent with chance noise properties and requires confirmation through future observations of this system. The sensitivity and spatial resolution are limited, and sufficient only to detect a very strong degree of non-axisymmetry. We can make a simple estimate of our ability to detect non-axisymmetry using the peak signal-to-noise ratio (SNR) per beam in the images: assuming that we want to detect non-axisymmetry at the 3$\sigma$ level, we would require beam-to-beam surface brightness changes of 100\% for disks with a peak SNR of 3-4, perhaps 50\% for disks with peak SNR of 5-8, or as little as 15-20\% for the SNR/beam of 20 reached by the ALMA observation of HD 107146. It is certainly possible that more sensitive future observations could detect a more subtle density contrast around the disk, especially for the lower-SNR detections in our sample. The lack of clumpy structure, even in systems with known giant planets, is consistent with theoretical work by \cite{Kuchner10}, demonstrating that collisions can smooth out structure even for millimeter grain sizes in disks with embedded planets.

 \section{Summary and Conclusions}

We have analyzed millimeter-wavelength interferometric observations from the SMA, CARMA, and ALMA of a sample of debris disks around Solar-type stars.  Two of the disks are spatially resolved for the first time by our observations.  We simultaneously model both the resolved millimeter-wavelength visibilities and the broadband SED of each system, which yields information about the geometry and basic dust grain properties in each disk. The inner radii of the debris belts tend to be a factor of a few larger than predicted from blackbody equilibrium calculations alone, implying that the disks contain small grains produced in a collisional cascade. The characteristic grain sizes derived from our modified blackbody approach are typically several times larger than the blowout size predicted for stars with Solar luminosities, consistent with results from previous studies.  Only one of the five disks, HD 107146, has a spatially resolved radial width ($\Delta R / R_\text{in} \gtrsim 1$).  We detect no asymmetries in the disks, to within the limits of our relatively low S/N ratio.  Overall, the five debris disks in our sample are consistent with scaled-up versions of the Solar system's Kuiper belt.  

\acknowledgments
The work of A.~S. was supported by a Student Observing Support grant from the National Radio Astronomy Observatory.  A.~M.~H. gratefully acknowledges support from NASA Origins of Solar Systems grant NNX13AI32G and NSF grant AST-1412647.  Support for CARMA construction was derived from the Gordon and Betty Moore Foundation, the Kenneth T. and Eileen L. Norris Foundation, the James S. McDonnell Foundation, the Associates of the California Institute of Technology, the University of Chicago, the states of California, Illinois, and Maryland, and the National Science Foundation.  CARMA development and operations were supported by the National Science Foundation under a cooperative agreement, and by the CARMA partner universities.  This paper makes use of the following ALMA data: ADS/JAO.ALMA\#2011.0.00470.S.  ALMA is a partnership of ESO (representing its member states), NSF (USA), and NINS (Japan), together with NRC (Canada) and NSC and ASIAA (Taiwan), in cooperation with the Republic of Chile.  The Joint ALMA Observatory is operated by ESO, AUI/NRAO and NAOJ.   The research made use of Astropy, a community-developed core Python package for Astronomy \citep{astropy13}.  

\newpage

\begin{thebibliography}{}
\expandafter\ifx\csname natexlab\endcsname\relax\def\natexlab#1{#1}\fi

\bibitem[{{Andrews} {et~al.}(2009){Andrews}, {Wilner}, {Hughes}, {Qi}, \&
  {Dullemond}}]{Andrews09}
{Andrews}, S.~M., {Wilner}, D.~J., {Hughes}, A.~M., {Qi}, C., \& {Dullemond},
  C.~P. 2009, \apj, 700, 1502

\bibitem[{{Apai} {et~al.}(2015){Apai}, {Schneider}, {Grady}, {Wyatt},
  {Lagrange}, {Kuchner}, {Stark}, \& {Lubow}}]{Apai15}
{Apai}, D., {Schneider}, G., {Grady}, C.~A., {et~al.} 2015, \apj, 800, 136

\bibitem[{{Ardila} {et~al.}(2004){Ardila}, {Golimowski}, {Krist}, {Clampin},
  {Williams}, {Blakeslee}, {Ford}, {Hartig}, \& {Illingworth}}]{Ardila04}
{Ardila}, D.~R., {Golimowski}, D.~A., {Krist}, J.~E., {et~al.} 2004, \apjl,
  617, L147

\bibitem[{{Astropy Collaboration} {et~al.}(2013){Astropy Collaboration},
  {Robitaille}, {Tollerud}, {Greenfield}, {Droettboom}, {Bray}, {Aldcroft},
  {Davis}, {Ginsburg}, {Price-Whelan}, {Kerzendorf}, {Conley}, {Crighton},
  {Barbary}, {Muna}, {Ferguson}, {Grollier}, {Parikh}, {Nair}, {Unther},
  {Deil}, {Woillez}, {Conseil}, {Kramer}, {Turner}, {Singer}, {Fox}, {Weaver},
  {Zabalza}, {Edwards}, {Azalee Bostroem}, {Burke}, {Casey}, {Crawford},
  {Dencheva}, {Ely}, {Jenness}, {Labrie}, {Lim}, {Pierfederici}, {Pontzen},
  {Ptak}, {Refsdal}, {Servillat}, \& {Streicher}}]{astropy13}
{Astropy Collaboration}, {Robitaille}, T.~P., {Tollerud}, E.~J., {et~al.} 2013,
  \aap, 558, A33

\bibitem[{{Backman} \& {Paresce}(1993)}]{Backman}
{Backman}, D.~E., \& {Paresce}, F. 1993, in Protostars and Planets III, ed.
  E.~H. {Levy} \& J.~I. {Lunine}, 1253--1304

\bibitem[{{Barucci} {et~al.}(2008){Barucci}, {Boehnhardt}, {Cruikshank}, \&
  {Morbidelli}}]{kuiper2009}
{Barucci}, M.~A., {Boehnhardt}, H., {Cruikshank}, D.~P., \& {Morbidelli}, A.
  2008, {The Solar System Beyond Neptune: Overview and Perspectives}, ed. M.~A.
  {Barucci}, H.~{Boehnhardt}, D.~P. {Cruikshank}, A.~{Morbidelli}, \&
  R.~{Dotson}, 3--10

\bibitem[{{Beckwith} \& {Sargent}(1991)}]{Beckwith}
{Beckwith}, S.~V.~W., \& {Sargent}, A.~I. 1991, \apj, 381, 250

\bibitem[{{Boley} {et~al.}(2012){Boley}, {Payne}, {Corder}, {Dent}, {Ford}, \&
  {Shabram}}]{Boley12}
{Boley}, A.~C., {Payne}, M.~J., {Corder}, S., {et~al.} 2012, \apjl, 750, L21

\bibitem[{{Booth} {et~al.}(2013){Booth}, {Kennedy}, {Sibthorpe}, {Matthews},
  {Wyatt}, {Duch{\^e}ne}, {Kavelaars}, {Rodriguez}, {Greaves}, {Koning},
  {Vican}, {Rieke}, {Su}, {Moro-Mart{\'{\i}}n}, \& {Kalas}}]{Booth}
{Booth}, M., {Kennedy}, G., {Sibthorpe}, B., {et~al.} 2013, \mnras, 428, 1263

\bibitem[{{Buenzli} {et~al.}(2010){Buenzli}, {Thalmann}, {Vigan}, {Boccaletti},
  {Chauvin}, {Augereau}, {Meyer}, {M{\'e}nard}, {Desidera}, {Messina},
  {Henning}, {Carson}, {Montagnier}, {Beuzit}, {Bonavita}, {Eggenberger},
  {Lagrange}, {Mesa}, {Mouillet}, \& {Quanz}}]{Buenzli10}
{Buenzli}, E., {Thalmann}, C., {Vigan}, A., {et~al.} 2010, \aap, 524, L1

\bibitem[{{Carpenter} {et~al.}(2008){Carpenter}, {Bouwman}, {Silverstone},
  {Kim}, {Stauffer}, {Cohen}, {Hines}, {Meyer}, \& {Crockett}}]{Carpenter08}
{Carpenter}, J.~M., {Bouwman}, J., {Silverstone}, M.~D., {et~al.} 2008, \apjs,
  179, 423

\bibitem[{{Chen} {et~al.}(2006){Chen}, {Sargent}, {Bohac}, {Kim},
  {Leibensperger}, {Jura}, {Najita}, {Forrest}, {Watson}, {Sloan}, \&
  {Keller}}]{Chen06}
{Chen}, C.~H., {Sargent}, B.~A., {Bohac}, C., {et~al.} 2006, \apjs, 166, 351

\bibitem[{{Chen} {et~al.}(2009){Chen}, {Sheehan}, {Watson}, {Manoj}, \&
  {Najita}}]{Chen09}
{Chen}, C.~H., {Sheehan}, P., {Watson}, D.~M., {et~al.} 2009, \apj, 701, 1367

\bibitem[{{Chen} {et~al.}(2014){Chen}, {Indebetouw}, {Muller}, {Kawamura},
  {Gordon}, {Sewi{\l}o}, {Whitney}, {Fukui}, {Madden}, {Meade}, {Meixner},
  {Oliveira}, {Robitaille}, {Seale}, {Shiao}, \& {van Loon}}]{Chen14}
{Chen}, C.-H.~R., {Indebetouw}, R., {Muller}, E., {et~al.} 2014, \apj, 785, 162

\bibitem[{{Chiang} {et~al.}(2009){Chiang}, {Kite}, {Kalas}, {Graham}, \&
  {Clampin}}]{Chiang09}
{Chiang}, E., {Kite}, E., {Kalas}, P., {Graham}, J.~R., \& {Clampin}, M. 2009,
  \apj, 693, 734

\bibitem[{{Chiang} {et~al.}(2003){Chiang}, {Jordan}, {Millis}, {Buie},
  {Wasserman}, {Elliot}, {Kern}, {Trilling}, {Meech}, \& {Wagner}}]{Chiang03}
{Chiang}, E.~I., {Jordan}, A.~B., {Millis}, R.~L., {et~al.} 2003, \aj, 126, 430

\bibitem[{{Corder} {et~al.}(2009){Corder}, {Carpenter}, {Sargent}, {Zauderer},
  {Wright}, {White}, {Woody}, {Teuben}, {Scott}, {Pound}, {Plambeck}, {Lamb},
  {Koda}, {Hodges}, {Hawkins}, \& {Bock}}]{Corder09}
{Corder}, S., {Carpenter}, J.~M., {Sargent}, A.~I., {et~al.} 2009, \apjl, 690,
  L65

\bibitem[{{Currie} {et~al.}(2011){Currie}, {Thalmann}, {Matsumura},
  {Madhusudhan}, {Burrows}, \& {Kuchner}}]{Currie11}
{Currie}, T., {Thalmann}, C., {Matsumura}, S., {et~al.} 2011, \apjl, 736, L33

\bibitem[{{Cutri} \& {et al.}(2012)}]{Cutri12}
{Cutri}, R.~M., \& {et al.} 2012, VizieR Online Data Catalog, 2311, 0

\bibitem[{{Debes} {et~al.}(2009){Debes}, {Weinberger}, \& {Kuchner}}]{Debes09}
{Debes}, J.~H., {Weinberger}, A.~J., \& {Kuchner}, M.~J. 2009, \apj, 702, 318

\bibitem[{{Dent} {et~al.}(2014){Dent}, {Wyatt}, {Roberge}, {Augereau},
  {Casassus}, {Corder}, {Greaves}, {de Gregorio-Monsalvo}, {Hales}, {Jackson},
  {Hughes}, {Lagrange}, {Matthews}, \& {Wilner}}]{Dent14}
{Dent}, W.~R.~F., {Wyatt}, M.~C., {Roberge}, A., {et~al.} 2014, Science, 343,
  1490

\bibitem[{{Dohnanyi}(1969)}]{dohnanyi69}
{Dohnanyi}, J.~S. 1969, \jgr, 74, 2531

\bibitem[{{Draine}(2006{\natexlab{a}})}]{Draine06}
{Draine}, B.~T. 2006{\natexlab{a}}, \apj, 636, 1114

\bibitem[{{Draine}(2006{\natexlab{b}})}]{Draine}
---. 2006{\natexlab{b}}, \apj, 636, 1114

\bibitem[{{Draine} \& {Lee}(1984)}]{DraineLee84}
{Draine}, B.~T., \& {Lee}, H.~M. 1984, \apj, 285, 89

\bibitem[{{Ertel} {et~al.}(2011){Ertel}, {Wolf}, {Metchev}, {Schneider},
  {Carpenter}, {Meyer}, {Hillenbrand}, \& {Silverstone}}]{Ertel11}
{Ertel}, S., {Wolf}, S., {Metchev}, S., {et~al.} 2011, \aap, 533, A132

\bibitem[{{Foreman-Mackey} {et~al.}(2013){Foreman-Mackey}, {Hogg}, {Lang}, \&
  {Goodman}}]{ForemanMackey}
{Foreman-Mackey}, D., {Hogg}, D.~W., {Lang}, D., \& {Goodman}, J. 2013, \pasp,
  125, 306

\bibitem[{{G{\'a}sp{\'a}r} {et~al.}(2012){G{\'a}sp{\'a}r}, {Psaltis}, {Rieke},
  \& {{\"O}zel}}]{gaspar11}
{G{\'a}sp{\'a}r}, A., {Psaltis}, D., {Rieke}, G.~H., \& {{\"O}zel}, F. 2012,
  \apj, 754, 74

\bibitem[{Goodman \& Weare(2010)}]{Goodman}
Goodman, J., \& Weare, J. 2010, Communications in Applied Mathematics and
  Computational Science, 5, 65

\bibitem[{{Hahn} \& {Malhotra}(2005)}]{Hahn05}
{Hahn}, J.~M., \& {Malhotra}, R. 2005, \aj, 130, 2392

\bibitem[{{Hillenbrand} {et~al.}(2008){Hillenbrand}, {Carpenter}, {Kim},
  {Meyer}, {Backman}, {Moro-Mart{\'{\i}}n}, {Hollenbach}, {Hines}, {Pascucci},
  \& {Bouwman}}]{Hillenbrand}
{Hillenbrand}, L.~A., {Carpenter}, J.~M., {Kim}, J.~S., {et~al.} 2008, \apj,
  677, 630

\bibitem[{{Hines} {et~al.}(2007){Hines}, {Schneider}, {Hollenbach}, {Mamajek},
  {Hillenbrand}, {Metchev}, {Meyer}, {Carpenter}, {Moro-Mart{\'{\i}}n},
  {Silverstone}, {Kim}, {Henning}, {Bouwman}, \& {Wolf}}]{Hines07}
{Hines}, D.~C., {Schneider}, G., {Hollenbach}, D., {et~al.} 2007, \apjl, 671,
  L165

\bibitem[{{Holland} {et~al.}(2003){Holland}, {Greaves}, {Dent}, {Wyatt},
  {Zuckerman}, {Webb}, {McCarthy}, {Coulson}, {Robson}, \& {Gear}}]{Holland03}
{Holland}, W.~S., {Greaves}, J.~S., {Dent}, W.~R.~F., {et~al.} 2003, \apj, 582,
  1141

\bibitem[{{Hughes} {et~al.}(2011){Hughes}, {Wilner}, {Andrews}, {Williams},
  {Su}, {Murray-Clay}, \& {Qi}}]{Hughes11}
{Hughes}, A.~M., {Wilner}, D.~J., {Andrews}, S.~M., {et~al.} 2011, \apj, 740,
  38

\bibitem[{{Hughes} {et~al.}(2012){Hughes}, {Wilner}, {Mason}, {Carpenter},
  {Plambeck}, {Chiang}, {Andrews}, {Williams}, {Hales}, {Su}, {Chiang},
  {Dicker}, {Korngut}, \& {Devlin}}]{Hughes12}
{Hughes}, A.~M., {Wilner}, D.~J., {Mason}, B., {et~al.} 2012, \apj, 750, 82

\bibitem[{{Janson} {et~al.}(2012){Janson}, {Carson}, {Lafreni{\`e}re},
  {Spiegel}, {Bent}, \& {Wong}}]{Janson12}
{Janson}, M., {Carson}, J.~C., {Lafreni{\`e}re}, D., {et~al.} 2012, \apj, 747,
  116

\bibitem[{{Kalas} {et~al.}(2013){Kalas}, {Graham}, {Fitzgerald}, \&
  {Clampin}}]{Kalas13}
{Kalas}, P., {Graham}, J.~R., {Fitzgerald}, M.~P., \& {Clampin}, M. 2013, \apj,
  775, 56

\bibitem[{{Kalas} {et~al.}(2004){Kalas}, {Liu}, \& {Matthews}}]{Kalas04}
{Kalas}, P., {Liu}, M.~C., \& {Matthews}, B.~C. 2004, Science, 303, 1990

\bibitem[{{Kennedy} \& {Wyatt}(2014)}]{Kennedy14}
{Kennedy}, G.~M., \& {Wyatt}, M.~C. 2014, \mnras, 444, 3164

\bibitem[{{Kharchenko} \& {Roeser}(2009)}]{Kharchenko09}
{Kharchenko}, N.~V., \& {Roeser}, S. 2009, VizieR Online Data Catalog, 1280, 0

\bibitem[{{Koerner} {et~al.}(2001){Koerner}, {Sargent}, \&
  {Ostroff}}]{Koerner01}
{Koerner}, D.~W., {Sargent}, A.~I., \& {Ostroff}, N.~A. 2001, \apjl, 560, L181

\bibitem[{{K{\'o}sp{\'a}l} {et~al.}(2013){K{\'o}sp{\'a}l}, {Mo{\'o}r},
  {Juh{\'a}sz}, {{\'A}brah{\'a}m}, {Apai}, {Csengeri}, {Grady}, {Henning},
  {Hughes}, {Kiss}, {Pascucci}, \& {Schmalzl}}]{Kospal13}
{K{\'o}sp{\'a}l}, {\'A}., {Mo{\'o}r}, A., {Juh{\'a}sz}, A., {et~al.} 2013,
  \apj, 776, 77

\bibitem[{{Krist} {et~al.}(2005){Krist}, {Ardila}, {Golimowski}, {Clampin},
  {Ford}, {Illingworth}, {Hartig}, {Bartko}, {Ben{\'{\i}}tez}, {Blakeslee},
  {Bouwens}, {Bradley}, {Broadhurst}, {Brown}, {Burrows}, {Cheng}, {Cross},
  {Demarco}, {Feldman}, {Franx}, {Goto}, {Gronwall}, {Holden}, {Homeier},
  {Infante}, {Kimble}, {Lesser}, {Martel}, {Mei}, {Menanteau}, {Meurer},
  {Miley}, {Motta}, {Postman}, {Rosati}, {Sirianni}, {Sparks}, {Tran},
  {Tsvetanov}, {White}, \& {Zheng}}]{Krist05}
{Krist}, J.~E., {Ardila}, D.~R., {Golimowski}, D.~A., {et~al.} 2005, \aj, 129,
  1008

\bibitem[{{Kuchner} \& {Stark}(2010)}]{Kuchner10}
{Kuchner}, M.~J., \& {Stark}, C.~C. 2010, \aj, 140, 1007

\bibitem[{{Lagrange} {et~al.}(2010){Lagrange}, {Bonnefoy}, {Chauvin}, {Apai},
  {Ehrenreich}, {Boccaletti}, {Gratadour}, {Rouan}, {Mouillet}, {Lacour}, \&
  {Kasper}}]{Lagrange10}
{Lagrange}, A.-M., {Bonnefoy}, M., {Chauvin}, G., {et~al.} 2010, Science, 329,
  57

\bibitem[{{Lagrange} {et~al.}(2012){Lagrange}, {Boccaletti}, {Milli},
  {Chauvin}, {Bonnefoy}, {Mouillet}, {Augereau}, {Girard}, {Lacour}, \&
  {Apai}}]{Lagrange12a}
{Lagrange}, A.-M., {Boccaletti}, A., {Milli}, J., {et~al.} 2012, \aap, 542, A40

\bibitem[{{Lejeune} {et~al.}(1997){Lejeune}, {Cuisinier}, \&
  {Buser}}]{Lejeune97}
{Lejeune}, T., {Cuisinier}, F., \& {Buser}, R. 1997, \aaps, 125, 229

\bibitem[{{Lyra} \& {Kuchner}(2013)}]{Lyra13}
{Lyra}, W., \& {Kuchner}, M. 2013, \nat, 499, 184

\bibitem[{{MacGregor} {et~al.}(2013){MacGregor}, {Wilner}, {Rosenfeld},
  {Andrews}, {Matthews}, {Hughes}, {Booth}, {Chiang}, {Graham}, {Kalas},
  {Kennedy}, \& {Sibthorpe}}]{MacGregor13}
{MacGregor}, M.~A., {Wilner}, D.~J., {Rosenfeld}, K.~A., {et~al.} 2013, \apjl,
  762, L21

\bibitem[{{Malmberg} \& {Davies}(2009)}]{Malmberg09}
{Malmberg}, D., \& {Davies}, M.~B. 2009, \mnras, 394, L26

\bibitem[{{Malmberg} {et~al.}(2011){Malmberg}, {Davies}, \&
  {Heggie}}]{Malmberg11}
{Malmberg}, D., {Davies}, M.~B., \& {Heggie}, D.~C. 2011, \mnras, 411, 859

\bibitem[{{Malmberg} {et~al.}(2007){Malmberg}, {de Angeli}, {Davies}, {Church},
  {Mackey}, \& {Wilkinson}}]{Malmberg07}
{Malmberg}, D., {de Angeli}, F., {Davies}, M.~B., {et~al.} 2007, \mnras, 378,
  1207

\bibitem[{{Maness} {et~al.}(2010){Maness}, {Kalas}, {Fitzgerald}, {Williams},
  {Chiang}, {Graham}, {Scherer}, {Peek}, {Hines}, {Schneider}, \&
  {Metchev}}]{Maness10}
{Maness}, H., {Kalas}, P., {Fitzgerald}, M., {et~al.} 2010, in Bulletin of the
  American Astronomical Society, Vol.~42, American Astronomical Society Meeting
  Abstracts 215, 361.04

\bibitem[{{Maness} {et~al.}(2008){Maness}, {Fitzgerald}, {Paladini}, {Kalas},
  {Duchene}, \& {Graham}}]{Maness08}
{Maness}, H.~L., {Fitzgerald}, M.~P., {Paladini}, R., {et~al.} 2008, \apjl,
  686, L25

\bibitem[{{Marois} {et~al.}(2008){Marois}, {Macintosh}, {Barman}, {Zuckerman},
  {Song}, {Patience}, {Lafreni{\`e}re}, \& {Doyon}}]{Marois08}
{Marois}, C., {Macintosh}, B., {Barman}, T., {et~al.} 2008, Science, 322, 1348

\bibitem[{{Marois} {et~al.}(2010){Marois}, {Zuckerman}, {Konopacky},
  {Macintosh}, \& {Barman}}]{Marois10}
{Marois}, C., {Zuckerman}, B., {Konopacky}, Q.~M., {Macintosh}, B., \&
  {Barman}, T. 2010, \nat, 468, 1080

\bibitem[{{Marzari} \& {Picogna}(2013)}]{Marzari13}
{Marzari}, F., \& {Picogna}, G. 2013, \aap, 550, A64

\bibitem[{{Matthews} {et~al.}(2014){Matthews}, {Krivov}, {Wyatt}, {Bryden}, \&
  {Eiroa}}]{Matthews14}
{Matthews}, B.~C., {Krivov}, A.~V., {Wyatt}, M.~C., {Bryden}, G., \& {Eiroa},
  C. 2014, ArXiv e-prints, arXiv:1401.0743

\bibitem[{{Mo{\'o}r} {et~al.}(2013){Mo{\'o}r}, {Juh{\'a}sz}, {K{\'o}sp{\'a}l},
  {{\'A}brah{\'a}m}, {Apai}, {Csengeri}, {Grady}, {Henning}, {Hughes}, {Kiss},
  {Pascucci}, {Schmalzl}, \& {Gab{\'a}nyi}}]{Moor13}
{Mo{\'o}r}, A., {Juh{\'a}sz}, A., {K{\'o}sp{\'a}l}, {\'A}., {et~al.} 2013,
  \apjl, 777, L25

\bibitem[{{Morales} {et~al.}(2013){Morales}, {Bryden}, {Werner}, \&
  {Stapelfeldt}}]{Morales13}
{Morales}, F.~Y., {Bryden}, G., {Werner}, M.~W., \& {Stapelfeldt}, K.~R. 2013,
  \apj, 776, 111

\bibitem[{{Morales} {et~al.}(2011){Morales}, {Rieke}, {Werner}, {Bryden},
  {Stapelfeldt}, \& {Su}}]{Morales11}
{Morales}, F.~Y., {Rieke}, G.~H., {Werner}, M.~W., {et~al.} 2011, \apjl, 730,
  L29

\bibitem[{{Moro-Martin}(2013)}]{MoroM13}
{Moro-Martin}, A. 2013, {Dusty Planetary Systems}, ed. T.~D. {Oswalt}, L.~M.
  {French}, \& P.~{Kalas}, 431

\bibitem[{{Mouillet} {et~al.}(1997){Mouillet}, {Larwood}, {Papaloizou}, \&
  {Lagrange}}]{Mouillet97}
{Mouillet}, D., {Larwood}, J.~D., {Papaloizou}, J.~C.~B., \& {Lagrange}, A.~M.
  1997, \mnras, 292, 896

\bibitem[{{Najita} \& {Williams}(2005)}]{NajWill}
{Najita}, J., \& {Williams}, J.~P. 2005, \apj, 635, 625

\bibitem[{{Pan} \& {Schlichting}(2012)}]{pan12}
{Pan}, M., \& {Schlichting}, H.~E. 2012, \apj, 747, 113

\bibitem[{{Patience} {et~al.}(2011){Patience}, {Bulger}, {King}, {Ayliffe},
  {Bate}, {Song}, {Pinte}, {Koda}, {Dowell}, \& {Kov{\'a}cs}}]{Patience11}
{Patience}, J., {Bulger}, J., {King}, R.~R., {et~al.} 2011, \aap, 531, L17

\bibitem[{{Pawellek} {et~al.}(2014){Pawellek}, {Krivov}, {Marshall},
  {Montesinos}, {{\'A}brah{\'a}m}, {Mo{\'o}r}, {Bryden}, \&
  {Eiroa}}]{Pawellek14}
{Pawellek}, N., {Krivov}, A.~V., {Marshall}, J.~P., {et~al.} 2014, ArXiv
  e-prints, arXiv:1407.4579

\bibitem[{{Pi{\'e}tu} {et~al.}(2011){Pi{\'e}tu}, {di Folco}, {Guilloteau},
  {Gueth}, \& {Cox}}]{Pietu11}
{Pi{\'e}tu}, V., {di Folco}, E., {Guilloteau}, S., {Gueth}, F., \& {Cox}, P.
  2011, \aap, 531, L2

\bibitem[{{Press} {et~al.}(2002){Press}, {Teukolsky}, {Vetterling}, \&
  {Flannery}}]{Recipes}
{Press}, W.~H., {Teukolsky}, S.~A., {Vetterling}, W.~T., \& {Flannery}, B.~P.
  2002, {Numerical recipes in C++ : the art of scientific computing}

\bibitem[{{Quillen}(2006)}]{Quillen06}
{Quillen}, A.~C. 2006, \mnras, 372, L14

\bibitem[{{Ricarte} {et~al.}(2013){Ricarte}, {Moldvai}, {Hughes},
  {Duch{\^e}ne}, {Williams}, {Andrews}, \& {Wilner}}]{Ricarte13}
{Ricarte}, A., {Moldvai}, N., {Hughes}, A.~M., {et~al.} 2013, \apj, 774, 80

\bibitem[{{Ricci} {et~al.}(2015){Ricci}, {Carpenter}, {Fu}, {Hughes}, {Corder},
  \& {Isella}}]{Ricci15}
{Ricci}, L., {Carpenter}, J.~M., {Fu}, B., {et~al.} 2015, \apj, 798, 124

\bibitem[{{Roccatagliata} {et~al.}(2009){Roccatagliata}, {Henning}, {Wolf},
  {Rodmann}, {Corder}, {Carpenter}, {Meyer}, \& {Dowell}}]{Rocca09}
{Roccatagliata}, V., {Henning}, T., {Wolf}, S., {et~al.} 2009, \aap, 497, 409

\bibitem[{{Schneider} {et~al.}(2014){Schneider}, {Grady}, {Hines}, {Stark},
  {Debes}, {Carson}, {Kuchner}, {Perrin}, {Weinberger}, {Wisniewski},
  {Silverstone}, {Jang-Condell}, {Henning}, {Woodgate}, {Serabyn},
  {Moro-Martin}, {Tamura}, {Hinz}, \& {Rodigas}}]{Schneider14}
{Schneider}, G., {Grady}, C.~A., {Hines}, D.~C., {et~al.} 2014, \aj, 148, 59

\bibitem[{{Smith} \& {Terrile}(1984)}]{Smith84}
{Smith}, B.~A., \& {Terrile}, R.~J. 1984, Science, 226, 1421

\bibitem[{{Soummer} {et~al.}(2014){Soummer}, {Perrin}, {Pueyo}, {Choquet},
  {Chen}, {Golimowski}, {Brendan Hagan}, {Mittal}, {Moerchen}, {N'Diaye},
  {Rajan}, {Wolff}, {Debes}, {Hines}, \& {Schneider}}]{Soummer14}
{Soummer}, R., {Perrin}, M.~D., {Pueyo}, L., {et~al.} 2014, \apjl, 786, L23

\bibitem[{{Strubbe} \& {Chiang}(2006)}]{Strubbe}
{Strubbe}, L.~E., \& {Chiang}, E.~I. 2006, \apj, 648, 652

\bibitem[{{Su} {et~al.}(2009){Su}, {Rieke}, {Stapelfeldt}, {Malhotra},
  {Bryden}, {Smith}, {Misselt}, {Moro-Martin}, \& {Williams}}]{Su09}
{Su}, K.~Y.~L., {Rieke}, G.~H., {Stapelfeldt}, K.~R., {et~al.} 2009, \apj, 705,
  314

\bibitem[{{Swift} {et~al.}(2013){Swift}, {Johnson}, {Morton}, {Crepp},
  {Montet}, {Fabrycky}, \& {Muirhead}}]{Swift}
{Swift}, J.~J., {Johnson}, J.~A., {Morton}, T.~D., {et~al.} 2013, \apj, 764,
  105

\bibitem[{{Williams} {et~al.}(2004){Williams}, {Najita}, {Liu}, {Bottinelli},
  {Carpenter}, {Hillenbrand}, {Meyer}, \& {Soderblom}}]{Williams04}
{Williams}, J.~P., {Najita}, J., {Liu}, M.~C., {et~al.} 2004, \apj, 604, 414

\bibitem[{{Wilner} {et~al.}(2011){Wilner}, {Andrews}, \& {Hughes}}]{Wilner11}
{Wilner}, D.~J., {Andrews}, S.~M., \& {Hughes}, A.~M. 2011, \apjl, 727, L42

\bibitem[{{Wilner} {et~al.}(2012){Wilner}, {Andrews}, {MacGregor}, \&
  {Hughes}}]{Wilner12}
{Wilner}, D.~J., {Andrews}, S.~M., {MacGregor}, M.~A., \& {Hughes}, A.~M. 2012,
  \apjl, 749, L27

\bibitem[{{Wilner} {et~al.}(2002){Wilner}, {Holman}, {Kuchner}, \&
  {Ho}}]{Wilner02}
{Wilner}, D.~J., {Holman}, M.~J., {Kuchner}, M.~J., \& {Ho}, P.~T.~P. 2002,
  \apjl, 569, L115

\bibitem[{{Wyatt}(2003)}]{Wyatt03}
{Wyatt}, M.~C. 2003, \apj, 598, 1321

\bibitem[{{Wyatt}(2006)}]{Wyatt06}
---. 2006, \apj, 639, 1153

\bibitem[{{Wyatt}(2008{\natexlab{a}})}]{Wyatt08}
---. 2008{\natexlab{a}}, ArXiv e-prints, arXiv:0807.1272

\bibitem[{{Wyatt}(2008{\natexlab{b}})}]{Wyatt08e}
---. 2008{\natexlab{b}}, \araa, 46, 339

\bibitem[{{Wyatt} \& {Dent}(2002)}]{Wyatt02f}
{Wyatt}, M.~C., \& {Dent}, W.~R.~F. 2002, \mnras, 334, 589

\bibitem[{{Zakamska} \& {Tremaine}(2004)}]{Zakamska04}
{Zakamska}, N.~L., \& {Tremaine}, S. 2004, \aj, 128, 869

\end{thebibliography}


 \end{document}